\documentclass[10pt]{article}

\newtheorem{lem}{Lemma}
\newtheorem{tm}{Theorem}

\newtheorem{de}{Definition}
\newtheorem{cl}{Collorary}

\begin{document}
\pagestyle{plain}
\begin{center}
\vskip10mm
   {\Large {\bf Structure behind Mechanics II: Deduction}}
\vskip7mm
\renewcommand{\thefootnote}{\dag}
{\large Toshihiko Ono}\footnote{ 
e-mail: BYQ02423@nifty.ne.jp \ \ or \ \ 
tono@swift.phys.s.u-tokyo.ac.jp}
\vskip5mm
\par\noindent
{\it 703 Shuwa Daiich Hachioji Residence,\\
4-2-7 Myojin-cho, Hachioji-shi, Tokyo 192-0046, Japan}
\end{center}
\vskip10mm
\setcounter{footnote}{0}
\renewcommand{\thefootnote}{\arabic{footnote}}

\begin{abstract}
This paper 
proves that protomechanics, previously introduced
in quant-ph/9909025,
 deduces both
 quantum mechanics
and classical mechanics.
It does not only solve the problem
of the arbitrariness on the
operator ordering for the quantization procedure,
but also that of the  analyticity at the exact classical-limit of $\hbar =0 $.
In addition, proto-mechanics 
proves valid also for the description of a half-spin.\\
\begin{center}
{\it To be Submitted to Found. Phys.}
\end{center}
\end{abstract}
\vskip10mm

\section{INTRODUCTION}

Previous paper \cite{SbMI}
proposed a basic theory on 
physical reality, 
named as {\it Structure behind Mechanics} (SbM).\footnote{Consult 
the letter \cite{PLA} on the overview of the present theory.}
It supposed that a field or a particle $X$
on the four-dimensional spacetime
has its internal-time $\tilde o_{{\cal P}(t)}(X)$ 
relative
to a domain ${\cal P}(t)$ 
of the four-dimensional spacetime,
whose boundary and interior
represent the present and the past at ordinary time $t\in {\bf R}$, 
respectively.
The classical action $S_{{\cal P}(t)}(X)$ realizes
internal-time $\tilde o_{{\cal P}(t)}(X) $ in the following relation:
\begin{equation}
\tilde o_{{\cal P}(t)}(X) = e^{iS_{{\cal P}(t)}(X)}.
\end{equation}
It further considered that
object $X$ also has the 
external-time $\tilde o_{{\cal P}(t)}^*(X)$ relative
to  ${{\cal P}(t)}$  which is the internal-time of all the rest but
$X$ in the universe.
Object $X$ gains the actual existence on  ${{\cal P}(t)}$
if 
and only if the internal-time
coincides with the external-time:
\begin{equation}\label{AX=XA}
\tilde o _{{\cal P}(t)}(X)= \tilde o_{{\cal P}(t)}^*(X) .
\end{equation}
This condition discretizes or quantizes the ordinary time passing from 
the past to the future,
and realizes the mathematical representation of
Whitehead's philosophy.
It also shows that 
object $X$ has its actual reality
only when it is related
with or
exposed to the rest of the world.
The both sides of relation (\ref{AX=XA}) further
obey the variational principle as
\begin{equation}
 \label{dAX=0/dXA=0}
\delta \tilde o_{{\cal P}(t)} (X)   = 0
\ \ \ , \ \ \ \
\delta \tilde o^*_{{\cal P}(t)} (X) = 0 .
\end{equation}  
These equations produce the equations of motion
in the deduced mechanics.

SbM
provided a foundation
for quantum mechanics and classical mechanics,
named as {\it protomechanics} \cite{SbMI},
originated by the past work \cite{Ono}.
The sapce $M$ of all the objects over present hypersurface 
$\partial {\cal P}(t)$
had an mapping $o_t : TM \to S^1$ for
the position $(x_t, \dot x_t )\in TM$ in the cotangent space
$TM$ 
corresponding to an object $X\in \tilde M$:
\begin{equation}\label{def eta0}
o_t \left( x_t,  \dot x_t \right) = 
\tilde o_{{\cal P}(t)} \left( X \right)  .
\end{equation}
For the velocity field $v_t \in X(M)$ 
such that $v_t \left( x_t\right) = 
 {{dx_t}\over {dt}} $, we will introduce a
section $\eta_t \in \Gamma \left[ E(M)\right] $
and call it {\it synchronicity} over $M$:
\begin{equation}\label{def eta}
\eta_t (x)=o_t \left( x, v_t (x) \right)   ;
\end{equation}
thereby, synchronicity $\eta_t$
has an 
information-theoretical sense,
as
defined for
the collective set of the objects $X$
that have different initial conditions
from one another.
On the other hand,
the emergence-frequency
$f_t\left( \eta_t \right) $  represent
the frequency that
object $ X$ satisfies condition (\ref{AX=XA}) on $ M$,
and
the true probability measure $\nu_t $ on $TM$
 representing the ignorance of the initial
position,
 defined the {\it emergence-measure}
 $\mu_t \left( \eta_t \right) $ as follows:
\begin{equation}
d\mu_t \left( \eta_t \right) (x)= d\nu_t \left( x, v_t(x) \right)  \cdot 
f_t
\left( \eta_t \right)  (x)  .
\end{equation}
The induced Hamiltonian $H^{T^*M}_t$ on $T^*M$,
further,
redefines the velocity field $v_t$
and the Lagrangian $L^{TM}_t$ as follows:
\begin{eqnarray}
v_t (x)
&=& {{\partial H^{T^*M}_t}\over {\partial p}}\left( x,
p\left( \eta_t \right) (x)
\right) \\
L^{TM}_t \left( x,
v(x)  \right)
&=&  v  (x) \cdot p \left(\eta _t \right) (x)   -  
H^{T^*M}_t\left( x, p\left(\eta _t \right) (x)  \right) ,
\end{eqnarray}
where mapping
$p$ satisfies
the modified Einstein-de Broglie relation:
\begin{equation}
p \left(\eta _t \right) =
- i \bar h   \eta_t  ^{ -1}
d \eta_t .
\end{equation}
The equation of motion
is the set of the following equations:
\begin{eqnarray}\label{EQOM}
\left( {{\partial }\over {\partial t}}
+{\cal L}_{v_t} \right) \eta_t (x)&=&  -i {\bar h}^{-1}
L_t^{TM}\left( x, v_t (x)\right)   \eta_t (x) ,\\
\left( {{\partial }\over {\partial t}}
+{\cal L}_{v_t} \right) d\mu_t \left( \eta_t \right) &=& 0.
\end{eqnarray}

Protomechanics had the statistical description on
an ensemble of all the
synchronicities $\eta_t^{\tau }$
for the labeling-time $\tau $
defined in the previous paper
such that $\eta_{\tau }^{\tau }= \eta$.
The next section
will be devoted to the review of such 
 statistical description for protomechanics.
Sections 3 and 4 will explain how  protomechanics
deduces classical mechanics and quantum mechanics,
respectively.
They will consider the space of the
synchronicities such that
\begin{equation}
\Gamma^A_{ k}  = \left\{ 
\eta \  \left\vert
\ \sup_{  U} p_j\left( \eta
\right) (x) = \hbar^{A} k_j \in {\bf R} \right.
\right\} 
\end{equation}
which requires $A=0 $ and $A=1 $ 
for
 classical case and quantum case, respectively.
Both cases will consider
a Lagrange foliation
$\bar p$ in $TM$ such that
it has a
synchronicity $\bar\eta [k]\in \Gamma^A_{ k} $
\begin{equation}
\bar p[k]=p\left( \bar  \eta [k] \right) ,
\end{equation}
and will separate every
synchronicity $\eta [k]\in \Gamma^A_{ k} $
into two parts:
\begin{equation}
\eta [k]= \bar  \eta [k] \cdot \xi .
\end{equation}
where $\xi \in \Gamma^A_{ 0} $.
Finally,
these sections will
compress all the infinite 
information
of back ground  $\xi $ to 
produce classical mechanics
and quantum mechanics.
Section 3 will additionally discuss
a consequent interpretation for the half-spin
of a particle;
a brief statement of
the conclusion will immediately follow.

Let me summarize the construction of the present paper
in the following diagram.

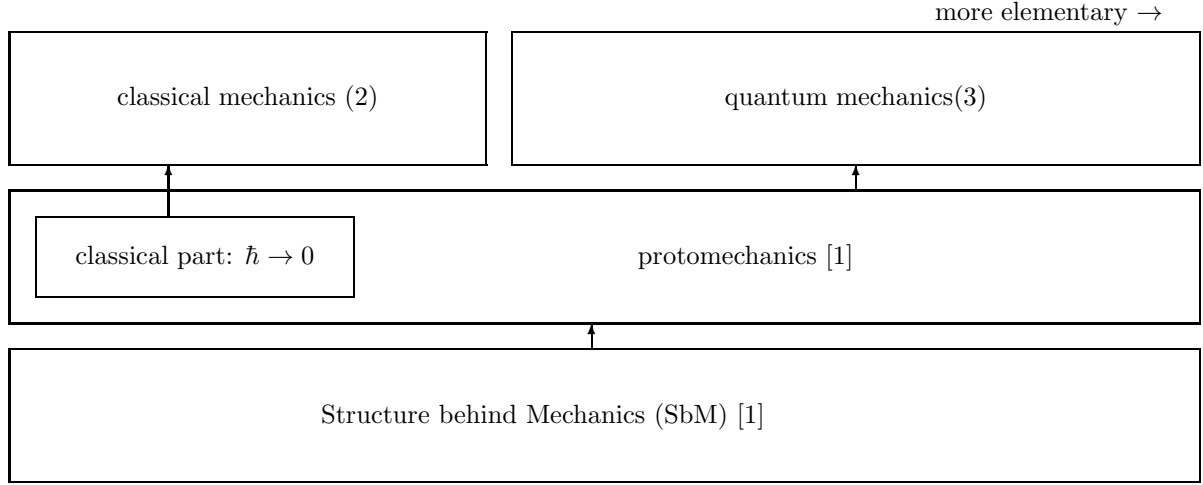
\begin{picture}(480,220)(15,20)

\put(10,160){\framebox(180,50){classical mechanics (2)}}

\put(200,160){\framebox(260,50){quantum mechanics(3)}}

\put(10,100){\framebox(450,50){\ \ \ \ \ \ \ \ \ \ \ \ \ \ \ \ \ \  
\ \ \ \ \ \ \ \ \ \ \ \ \ \  protomechanics \cite{SbMI}}}
\put(20,110){\framebox(120,30){classical part: $\hbar \to 0$}}

\put(10,40){\framebox(450,50){Structure behind Mechanics (SbM) \cite{SbMI} \ \ \ \ \ \ \ \ \ \ \ \ \ }}

\put(230,90){\vector(0,1){10}}
\put(70,140){\vector(0,1){20}}
\put(330,150){\vector(0,1){10}}

\put(360,215){more elementary $ \rightarrow $}
\put(20,20){$ \downarrow $ more fundamental}
\put(250,20){{\small * Numbers in bracket $( \ )$ refer those of sections.}}
\end{picture}
{}\\
{}\\

In this paper, 
$c$ and
 $h$
denote the speed of light and Planck's constant, respectively.
I will
use Einstein's rule
in the tensor calculus
for Roman indices'
$i, j, k \in {\bf N}^N$
and Greek indices' 
$\nu , \mu  \in {\bf N}^N$,
and not for Greek indices' 
$\alpha , \beta , \gamma \in {\bf N}^N$,
and I
further denote the trace (or supertrace) operation
of a quantum observable $\hat F$ as
$ \langle \hat F  \rangle $
that is only one difference 
from the ordinary notations in quantum mechanics,
where $i=\sqrt{-1}$.

\section{Review on Protomechanics}

Let us review
the protomechanics 
in the statistical way
for the ensemble of all the synchronicities on $M$,
and
construct
the dynamical description
for the 
collective motion
of the sections of $E(M)$.
Such statistical description realizes
the description within a
long-time interval 
through the introduced relabeling process
so as to change the labeling time,
that is the time for the
initial condition
before analytical problems occur.
In addition,
it clarifies the relationship between
classical mechanics
and quantum mechanics
under the assumption
that the present theory safely
induces them,
and that will be proved in the following sections.\footnote{
In another way,
consult quant-ph/9906130.}
For mathematical simplicity,
the discussion below
suppose that
$M$ is a $N-$dimensional manifold for a finite 
natural number
$N\in {\bf N}$.

The derivative operator $D= \hbar dx^{j}\partial_{j} :
 T_0^m (M) \to T_0^{m+1} (M)$
($m\in {\bf N}$)
for the space $T^n_0 (M)$ of all the $(0,n)$-tensors on $M$
can be described as
\begin{equation}\label{D-der}
 D^n p(x) = \hbar ^n \left( \prod_{k=1}^{n}\partial_{j_k} p_j(x) \right)
dx^j \otimes \left( \otimes_{k =1}^{n} dx^{j_k} \right) .
\end{equation}
By utilizing this derivative operator $D$,
the following norm for every $p\in \Lambda^1 (M)$
endows space $\Lambda^1 (M)$ 
 with a norm topology:
\begin{equation}\label{q-norm}
\left\|   p \right\| = \sup_{M} 
\sum_{\kappa \in {\bf Z}_{\geq 0} } 
\left\vert D^{\kappa } 
p(x) \right\vert_x ,
\end{equation} 
where $\vert \ \ \vert_x $ is a norm of covectors at $x \in M$.
In terms of this norm topology,
we can consider the space
$C^{\infty } \left( \Lambda^1 \left( M \right)   , C^{\infty }(M) \right) $
of all the $C^{\infty }$-differentiable
mapping from $\Lambda^1 \left( M \right) $ to $C^{\infty }(M)=C^{\infty }(M, {\bf R})$
and the subspaces
of the space
$ 
C( \Gamma  [
E ( M  )  ]   ) $ such that
\begin{equation}
C\left( \Gamma \left[
E\left( M \right) \right]  \right) 
= \left\{  \left. p^*F : 
\Gamma \left[
E ( M ) \right] \to C^{\infty }(M) \  \right\vert  
  F \in   
C^{\infty }\left( \Lambda^1  ( M  ) , C^{\infty }(M) \right)
 \right\} .
\end{equation}
Classical mechanics
requires the local 
dependence on
the momentum for functionals,
while quantum mechanics
needs the wider class
of functions
that depends on their derivatives.
The space of the classical
functionals
and
that of the quantum functionals
are defined as
\begin{eqnarray}
C_{cl}\left( \Gamma \left[
E\left( M \right) \right] \right) 
&=& \left\{ p^*F \in C\left( \Gamma \left[
E\left( M \right) \right] \right)
\ \left\vert \
 p^*F  \left( \eta \right) (x) = 
{\bf F }\left( x ,    p (\eta ) (x)  \right) \
\right.
\right\} \\
C_{q\ }\left( \Gamma \left[
E\left( M \right) \right]  \right) 
&=& \left\{   p^*F \in C \left( \Gamma \left[
E\left( M \right) \right] \right)
\ \right\vert \\
& \ & \ \ \ \ \ \ \ \ \ \ \left.
 p^*F  \left( \eta \right) (x) = 
{\bf F }\left( x , p (\eta ) (x),  
..., D^{n} p (\eta ) (x) , ... \right) \ 
\right\} ,
\end{eqnarray}
and related with each other as
\begin{equation}\label{increasing}
C_{cl}\left( \Gamma \left[
E\left( M \right) \right]  \right) 
\subset
C_q\left( \Gamma \left[
E\left( M \right) \right]  \right) 
\subset
C\left( \Gamma \left[
E\left( M \right) \right] \right) .
\end{equation}
In other words,
the classical-limit
indicates
the limit of $\hbar \to 0$ with fixing $\vert p(\eta )(x) \vert $
finite at every $x\in M$, or what
the characteristic length $[x]$ and momentum $[p]$
such that $x/[x] \approx 1 $ and $p/[p] \approx 1 $
satisfies 
\begin{equation}
[p]^{-n-1} D^n p(\eta )(x)  \ll 1 .
\end{equation}

On the other hand,
the emergence-measure $\mu (\eta )  $ has
the Radon measure 
$ \tilde \mu (\eta )  $
for section $\eta \in \Gamma [E(M)]$ such that
\begin{equation}
\tilde \mu (\eta ) \ \left( p^*F \left(  \eta  \right)
  \right)  =
\int_M d\mu (\eta ) (x) 
p^*F \left( \eta   \right) (x) .
\end{equation}
Let
us assume
set $ \Gamma \left( E(M) \right) $
is a measure space
having the probability 
measure ${\cal M}$ such that 
\begin{equation}
 {\cal M} \left(   
\Gamma \left( E(M) \right)  \right) =1.
\end{equation}
For a subset $C_n\left( \Gamma \left( E(M) \right)   \right)
\subset C\left( \Gamma \left( E(M) \right)   \right) $,
 an element
$\bar \mu 
\in C_n\left( \Gamma \left( E(M) \right)   \right) ^* $
is a linear functional
 $ \bar \mu : C_n\left( 
\Gamma \left[ E(M) \right] \right) 
 \to {\bf R}   $
such that
\begin{eqnarray}
 \bar  \mu \left( p^* F \right) &=&
\int_{\Gamma \left[ E(M) \right] }d {\cal M} (\eta ) \
\tilde  \mu (\eta ) \ \left( p^* F \left(  \eta   \right)
  \right)  \\
&=& \int_{\Gamma \left[ E(M) \right]}d {\cal M} (\eta ) \
\int_M dv(x) \
\rho \left(\eta \right) (x)
F  \left( p ( \eta  ) \right) (x)   ,
\end{eqnarray}
where $ d \mu (\eta )   =
dv  \
\rho \left(\eta \right)   $.
Let us call mapping $\rho :\Gamma [E(M)]\to C^{\infty }(M)$
as the {\it emergence-density}.
The dual spaces
make an decreasing series
of subsets:
\begin{equation}\label{decreasing}
C_{cl}\left( \Gamma \left( E(M) \right)   \right) ^*
\supset
C_q\left( \Gamma \left( E(M) \right)   \right) ^*
\supset
C \left( \Gamma \left( E(M) \right)   \right) ^*.
\end{equation}
Let us summarize
how the relation between quantum mechanics and
classical mechanics in the following
diagram.

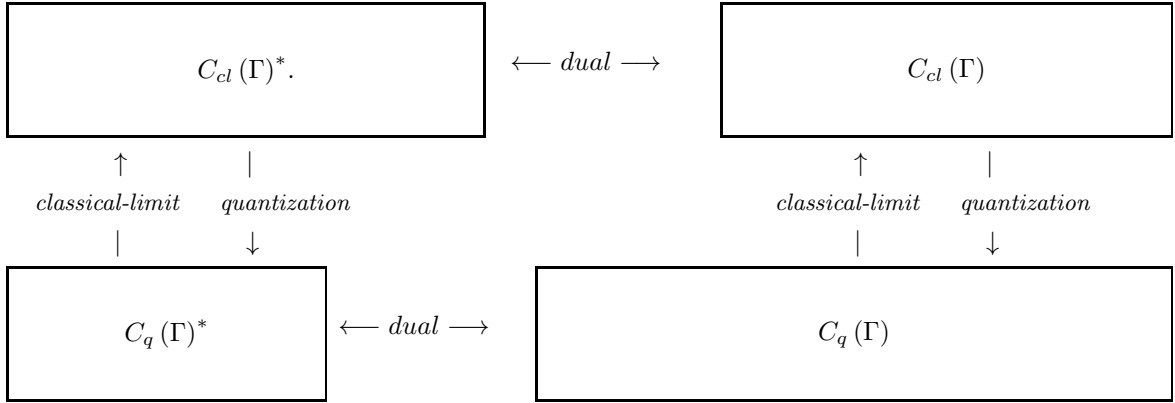
\begin{picture}(380,220)(10,20)
\put(210,50){\framebox(240,50){$C_{q}\left( \Gamma \right) $}}
\put(280,150){\framebox(170,50){$C_{cl}\left( \Gamma \right) $}}
\put(10,50){\framebox(120,50){$C_{q}\left( \Gamma \right) ^*$}}
\put(10,150){\framebox(180,50){$C_{cl}\left( \Gamma \right) ^*$.}}
\put(135,75){$ \longleftarrow   
dual  \longrightarrow $}
\put(200,175){$ \longleftarrow  
dual  \longrightarrow $}
\put(50,137){$ \uparrow $}
\put(20,122){{\small\it classical-limit}}
\put(50,107){$ \vert $}
\put(330,137){$ \uparrow $}
\put(300,122){{\small\it classical-limit}}
\put(330,107){$ \vert $}
\put(100,137){$ \vert $}
\put(90,122){{\small\it quantization}}
\put(100,107){$ \downarrow $}
\put(380,137){$ \vert $}
\put(370,122){{\small\it quantization}}
\put(380,107){$ \downarrow $}
\end{picture}

To investigate the time-development
of the statistical state discussed so far,
we will introduce the related group.
The group ${\cal D}(M)$ of
all the
$C^{\infty }$-diffeomorphisms of $M$
and the abelian group
$ C^{\infty }\left( 
M \right) $
 of all the $C^{\infty }$-functions 
on $M$
construct the semidirect product
$ S  ( M ) =
{\cal D} (M) \times_{semi. } C^{\infty }(M) $
of ${\cal D} (M)$ with $ C^{\infty }(M) $,
and define
the multiplication $\cdot $
between $ \Phi_1=(\varphi_1,s_1)$ and $\Phi_2=(\varphi_2,s_2)\in   S  ( M  )
$ as  
\begin{equation}\label{(2.1.1)}
\Phi_1\cdot \Phi_2 =(\varphi _1 \circ \varphi _2,(\varphi_2^*
s_1 )\cdot s_2) ,  
\end{equation}
for the pullback $\varphi ^*$ by $\varphi \in
{\cal D} (M)$.
The
Lie algebra 
$s (M)$
of $ S (M) $ 
has the Lie bracket such that, for   
$ V_1=(v_1 , U_1)$ and $V_2=(v_2 , U_2) \in  s (M)$,
\begin{equation}\label{(2.1.2)}
 [V_1, V_2 ]= \left( [v_1 ,v_2 ], v_1 U_2 - v_2 U_1 
+ \left[ U_1 , U_2 \right] \right) ;  
\end{equation}
 and its dual space $s (M)^*$
is defined by natural pairing $\langle \ , \  \rangle $.
Lie group
$ S  ( M  )  $ now acts on every $C^{\infty }$ section
 of $E ( M   ) $ (consult {\it APPENDIX}).
We shall further introduce the group $Q(M) 
= Map\left( \Gamma \left[ E(M)\right] , S  (M) \right) $ 
of all the mapping from $ \Gamma \left[ E(M)\right] $ into $S(M) $,
that has
the Lie algebra $q(M) 
= Map\left(  \Gamma \left[ E(M)\right] , s  (M)\right) $ and
its dual space $ 
q(M)^* = Map \left(   \Gamma \left[ E(M)\right] , s  (M)^* \right)  $.

Let us consider the time-development
of the section $\eta ^{\tau }_{t }(\eta ) \in \Gamma [E(M)]$
such that
 the {\it labeling time} $\tau $ satisfies
$\eta ^{\tau }_{\tau }(\eta ) =\eta  $.
It has the momentum
 $p_t^{\tau }(\eta )
=-i \bar h\eta ^{\tau }_t(\eta )^{-1}
d \eta ^{\tau }_t(\eta ) $ and
the emergence-measure $\mu^{\tau }_t(\eta ) $
such that
\begin{equation}\label{q-measure-rel}
d {\cal M} \left( \eta \right) \
  \tilde \mu_t^{\tau } \left( \eta  \right) 
= d {\cal M} \left(  \eta ^{\tau }_t(\eta )\right) \
\tilde \mu_t \left( \eta ^{\tau }_t(\eta )\right) :
\end{equation}
\begin{eqnarray}\label{mu(F)}
\bar  \mu_t \left( p^*F_t\right) 
&=&\int_{\Gamma \left[ E(M) \right]}d {\cal M} (\eta ) \
 \tilde \mu_t (\eta ) \ \left( p^*F_t  (\eta )
  \right)  \\
&=& \int_{\Gamma \left[ E(M) \right]}d {\cal M} \left( \eta \right) \
\tilde \mu^{\tau }_t \left( \eta  \right) \ \left( p^*F  \left( 
\eta ^{\tau }_t(\eta )\right) 
  \right)  \\
&=&
 \int_{\Gamma \left[ E(M) \right]}d {\cal M} \left( \eta \right) \
\int_M dv(x) \ \rho_t^{\tau } (\eta ) (x) 
F_t   \left( p ^{\tau }_t(\eta )\right)  (x)    .
\end{eqnarray}
The introduced labeling time $\tau $
can always be chosen such that $ \eta ^{\tau }_t(\eta ) $
does not have any singularity
within a short time for every $\eta
\in \Gamma \left[ E(M) \right]$.
The emergence-momentum
$ {\cal J}_t^{\tau }  
\in   q\left(M\right) ^*$ 
such that
\begin{eqnarray}
{\cal J}_t^{\tau }  (\eta ) &=&
  d {\cal M} \left( \eta ^{\tau }_t(\eta )\right) \
\left( \tilde   \mu_t \left( \eta ^{\tau }_t(\eta )\right)  
 \otimes 
p_t^{\tau }(\eta )  ,\tilde   \mu_t \left( \eta ^{\tau }_t(\eta )\right)  \right) \\
&=&  d {\cal M} (\eta ) \
 \left(  \tilde \mu_t ^{\tau }
\left( \eta  \right) \otimes 
p_t^{\tau }(\eta )  ,
\tilde \mu_t ^{\tau }
\left( \eta  \right) \right) \ 
\end{eqnarray}
satisfies the following relation
for the functional
${\cal F}_t   : q\left(M\right) ^* \to {\bf R} $:
\begin{equation}
{\cal F}_t  \left(  {\cal J}^{\tau }_t\right)
=\bar \mu_t \left( p^* F_t\right) ,
\end{equation}
whose value is independent of
labeling time $\tau $.
The operator $\hat F_t^{\tau } =
{{\partial {\cal F}_t}
\over {\partial {\cal J}}}  \left( {\cal J}^{\tau }_t\right) $ 
is defined as
\begin{equation}
\left. {d\over d\epsilon }\right\vert_{\epsilon =0}
{\cal F }_t  \left( {\cal J}^{\tau }_t + \epsilon {\cal K}\right) 
= \left\langle  {\cal K} , \hat F_t^{\tau }  \right\rangle ,
\end{equation}
i.e.,
\begin{equation}
\hat F_t  ^{\tau }
= \left(   {\cal D}_{ \rho_t^{\tau }(\eta ) } F_t  
\left(  p^{\tau }_t (\eta ) \right) ,
- p_t^{\tau } (\eta )  \cdot     {\cal D}_{ \rho_t^{\tau }(\eta ) } F_t  
\left(  p^{\tau }_t (\eta ) \right)
+ F_t   \left( p^{\tau }_t (\eta )  \right)
\right)  ,
\end{equation}
where the derivative
$  {\cal D}_{\rho  } 
F   \left( p \right)   $ can be 
introduced as follows
excepting the point
where the distribution $\rho $
becomes zero:
\begin{equation}
{\cal D}_{\rho  } 
F   \left( p \right)  (x)
=\sum_{ (n_1 ,  ... , n_N) \in {\bf N}^N}
{1\over { \rho  (x)} }
\left\{ \prod_{i}^{N} \left( - \partial_i \right) ^{ n_i }
  \left(  \rho  (x) 
p ( x )
{{\partial F    }\over { \partial \left\{
\left(
 \prod_{i}^{N}
\partial_i^{n_i } \right) p_j \right\}  }}
\right)  \right\}
\partial_j .
\end{equation}
Thus, the following null-lagrangian relation
 can be obtained:
\begin{equation}
{\cal F}_t   
 \left( {\cal J}_t^{\tau }  \right) 
= 
\langle {\cal J}_t^{\tau }  , \hat F _t ^{\tau } \rangle  ,
\end{equation}
while
the normalization condition 
has the following expression:
\begin{equation}\label{q-normalize}
{\cal I}\left( {\cal J}^{\tau }_t \right) =1 
\ \ \ \ \ for \ \ \ \ \ 
{\cal I}\left( {\cal J}^{\tau }_t \right) =
\int_{\Gamma \left[ E(M) \right] }d{\cal M}(\eta )  \
\mu_t (\eta ) (M) .
\end{equation}

For Hamiltonian operator $\hat H_t^{\tau }  =
{{\partial {\cal H}_t}\over {\partial {\cal J}}} 
  \left( {\cal J}_t^{\tau } \right)\in   q\left(  M\right) $
corresponding to Hamiltonian
$p^*H_t \left( \eta \right) (x) =
H^{T^*M}_t\left( x, p \left( \eta \right) \right) $,
equations (\ref{EQOM}) of motion 
becomes Lie-Poisson equation 
\begin{equation}\label{Hq}
{{\partial  {\cal J} ^{\tau }_t}\over {\partial t}} 
= ad^*_{\hat H_t ^{\tau }     }{\cal J}^{\tau } _t   ,
\end{equation}
which can be expressed as
\begin{equation}
{{\partial  } \over {\partial t}} 
 \rho _t^{\tau }(\eta )(x) 
= -\surd ^{-1}\partial_j \left( 
{ { \partial H_t^{T^*M}  }\over {\partial \ p_j \  }} 
\left( x, p_t^{\tau } \left( \eta \right) (x) \right) 
 \rho  _t^{\tau }(\eta )(x) \surd  \right) ,
\end{equation}
\begin{eqnarray}\nonumber
{{\partial  } \over {\partial t}}
\left( \rho  _t^{\tau }(\eta )(x) 
  p_{tk}^{\tau } (\eta ) (x) \right)
\nonumber
&=&
 - \surd^{-1} \partial_j \left(
{ { \partial H_t^{T^*M}  }\over {\partial \ p_j \  }} 
\left( x, p_t^{\tau } \left( \eta \right) (x) \right) 
\rho  _t^{\tau }(\eta )(x)  p_{tk}^{\tau } (\eta )(x) \surd  
\right) \\
\nonumber
&  &
-  \rho  _t^{\tau }(\eta )(x)  p_{tj}^{\tau } (\eta ) (x) 
\partial_k  
{ { \partial H_t^{T^*M}  }\over {\partial \ p_j \  }} 
\left( x, p_t^{\tau } \left( \eta \right) (x) \right)   \\
&  &
+ \rho  _t^{\tau }(\eta ) (x)  
\partial_k \left(
 p^{\tau }_t (\eta ) (x) \cdot 
{ { \partial H_t^{T^*M}  }\over {\partial \ p \  }} 
\left( x, p_t^{\tau } \left( \eta \right) (x) \right) 
\right. \\
&  & \ \ \ \ \ \ \ \ \ \ \ \ \ \ \ \ \
\left.
- H_t^{T^*M}   \left( x, p^{\tau }_t (\eta ) (x) \right) 
\right)  .
\end{eqnarray}
Equation (\ref{Hq}) will
prove in the following sections 
to include the
Schr\"odinger equation 
in canonical quantum mechanics
and the classical Liouville equations
in classical mechanics.

For
$  {\cal U}_t  ^{\tau }
 \in  Q\left(  M\right) $ such that
$
{{\partial  
{\cal U}_t  ^{\tau } }\over {\partial t}}  
\circ \left( {\cal U}_t  ^{\tau  } \right) ^{-1}= 
\hat H_t  ^{\tau } (\eta ) 
\in q(M)  $,
let us introduce the following operators:
\begin{eqnarray}
  \tilde H_t^{\tau } (\eta )   = Ad^{-1}_{{\cal U}_t  ^{\tau }   }
\hat H_t^{\tau } (\eta )  \ \ \left( =  \hat H_t ^{\tau } (\eta )\right) , \   and \ \ \
 \tilde F_t^{\tau } (\eta )  = Ad^{-1}_{{\cal U}_t  ^{\tau }    }
\hat F_t ^{\tau } (\eta ) .
\end{eqnarray}
Lie-Poisson equation (\ref{Hq})
is
equivalent to the following equation:
\begin{equation}\label{general heisenberg}
{{\partial }\over {\partial t}}\tilde F_t  ^{\tau }    = 
\left[ \tilde H _t ^{\tau }    ,\tilde F_t   ^{\tau }   \right] 
+\widetilde{\left( {{\partial F_t  ^{\tau }     }
\over {\partial t}} \right) } .
\end{equation}
The general theory
for Lie-Poisson systems
certificates that,
 if a  group action of Lie group
$ Q(M)$
keeps the Hamiltonian ${\cal H}_t: q(M)^* \to {\bf R}$  
invariant,
there exists an invariant charge 
functional
$Q : \Gamma \left[ E(M)\right] \to
C(M) $ and the induced
function ${\cal Q}: q(M)^* \to {\bf R}$ such that
\begin{equation}\label{charge-invariant}
\left[ \hat H ^{\tau }_t, \hat Q^{\tau } \right]  = 0 .
\end{equation}

\section{DEDUCTION OF CLASSICAL MECHANICS}

In classical Hamiltonian mechanics,
the  state of a particle 
on  manifold $M$
can be
represented as a position  in
the cotangent bundle $T^*M$.
In this section,  we will reproduce
the classical equation of motion
from the general theory
presented in the previous section.
Let us here concentrate ourselves
on the case where $M$ is $N$-dimensional
manifold for simplicity,
though the discussion
below would still be valid if
substituting an appropriate Hilbert space
when
$M$ is infinite-dimensional
ILH-manifold\cite{Omori}.

\subsection{Description of Statistical State}

Now,
we must be concentrated
on the case
where the physical
 functional $F\in C^{\infty }\left( \Lambda^1(M) , C^{\infty }(M) \right) $
does {\it not}
depend  on the derivatives 
of the $C^{\infty }$ 1-form $p\left( 
\eta \right)\in \Lambda^1 (M) $
induced from $\eta \in \Gamma \left[ E(M) \right] $, then it
has the following expression:
\begin{equation}\label{local F}
p^* F  \left( \eta \right) (x) = 
F^{T^*M} \left( x,  p\left( \eta \right) (x) \right)   .
\end{equation}

Let us  choose a  
coordinate system  
$\left( U_{\alpha } , {\bf x}_{\alpha  } 
\right) _{\alpha \in \Lambda_M}$
for  a covering $\left\{ U_{\alpha } 
\right\} _{\alpha \in \Lambda _{M  }} $
over $M$,
i.e.,
$ M = \bigcup_{\alpha \in \Lambda_M }
U_{\alpha }$.
Let us further choose
a reference set $U \subset U_{\alpha }$
such that $v(U) \neq 0$
and consider the set $\Gamma_{U k}\left[ E(M)\right] $ of the
$C^{\infty }$ sections of  $E(M)$
having corresponding momentum
$p\left( \eta \right) $
the
supremum of whose every component
$p_j\left( \eta \right) $ 
in $U$ becomes
 the value $k_j   $ for $ k= (k_1 ,..., k_N) \in {\bf R}^N$:\footnote{
To substitute
$\Gamma_{U k}\left[ E(M)\right] = \left\{ 
\eta \in \Gamma \left[ E(M)\right] \  \left\vert
\ \int_{  U}dv(x) \ p_j\left( \eta
\right) (x) = k_j v\left( U \right) \right.
\right\} $ for
definition (\ref{lambda cl.})
also induces the similar discussion
below,
while there exist a variety of the classification
methods that produce the same result.}
\begin{equation}\label{lambda cl.}
\Gamma_{U k} \left[ E(M)\right] = \left\{ 
\eta \in \Gamma \left[ E(M)\right] \  \left\vert
\ \sup_{  U} p_j\left( \eta
\right) (x) =  k_j \right.
\right\} .
\end{equation}
Thus,
every section $\eta \in \Gamma\left[ E(M)\right] $ has 
some $k\in {\bf R}^N$
such that $\eta =\eta [k]\in \Gamma_{U k} \left[ E(M)\right] $.
Notice that
$\Gamma_{U k} \left[ E(M)\right] $
can be identified with
$\Gamma_{U^{\prime } k} \left[ E(M)\right] $
for every two reference sets $U$ and $U^{\prime }\in M$,
since
there
exists a diffeomorphism $\varphi $
satisfying $\varphi \left( U \right) =U^{\prime } $;
thereby, we will simply denote
$\Gamma_{U k}\left[ E(M)\right] $
as $\Gamma_{ k}\left[ E(M)\right] $.

On the other hand,
let us consider
the space $L\left( T^*M \right) $ of
all the Lagrange foliations,
i.e.,
every element
$\bar p \in L\left( T^*M \right) $
is a mapping $\bar  p :
{\bf R}^N \to \Lambda^1(M) $
such that
each $q  \in T^*M$
has a unique $k \in {\bf R}^N $  as
\begin{equation}
q =\bar  p [ k ] \left( \pi (q) \right) .
\end{equation}
For every $\bar  p 
= p\circ \bar  \eta  \in L\left( T^*M \right) $
such that $\bar  \eta [k] \in \Gamma_{  k}\left[ E(M)\right]  $,
it is possible to
separate an element $\eta [k] \in \Gamma_{  k}\left[ E(M)\right] $
for a
$\xi\in  \Gamma_{ 0}\left[ E(M)\right]  $ as
\begin{equation}
\eta [k]= \bar  \eta [k] \cdot \xi ,
\end{equation}
or to 
separate 
momentum
$p \left( \eta [ k] \right) $ as
\begin{equation}\label{cl-separation}
p \left( \eta [ k] \right) 
= \bar  p    [ k]  
  + p  \left( \xi \right) ;
\end{equation}
thereby, we can express 
the emergence-density $ \rho  
: \Gamma [E(M)] \to C^{\infty }\left( M\right) $ 
in the following form
for the  function $ \varrho \left(  \xi \right)  \in  C^{\infty }
\left( T^*M,{\bf R}\right) $ on $T^*M$:
\begin{equation}\label{rho=rho}
 \rho    \left( \eta [k] \right) 
(x) \surd 
=  
\varrho \left(  \xi \right) \left( x, p \left( \eta [k]\right) (x) \right) .
\end{equation}

We call the set
$ B\left[ E(M)\right]  
= \Gamma_{0}\left[ E(M)\right]  $
the {\bf back ground} of $L\left( T^*M \right) $.
For the Jacobian-determinant
$ \sigma  [ k]  = det 
\left( {{\partial \bar p_{ti} ^{\tau }  [ k]   }
\over {\partial k_j}}\right) $,
we will define the measure
${\cal N}$ on $B\left[ E(M)\right] $
for the $\sigma $-algebra induced from that of $\Gamma \left[
E(M)\right] $:
\begin{equation}
d {\cal M} \left( \eta [k]\right) \
d v (x)
= d^Nk  d {\cal N}  \left(  \xi \right) 
 d v (x) \ \sigma [k](x)\  .
\end{equation}
For  separation (\ref{cl-separation}),
the Radon measure $\tilde  \mu (\eta )$
 induces the measure
$\omega^N $ on $T^*M$
in
the following lemma
such that $\omega^N = \phi_{U_{\alpha }
* } d^Nx \wedge d^Nk$
for
$d^Nx =dx^1 \wedge ... dx^N $
and $d^Nk =dk^1 \wedge ... dk^N $.
\begin{lem}
The following relation holds:
\begin{equation}
\bar \mu \left( p^*F\right)=
\int_{T^*M} \omega^N (q) \
\rho^{T^*M} \left( q \right)  
F^{T^*M} \left( q \right)  ,
\end{equation}
where
\begin{equation}
\rho^{T^*M} \left( q \right)  =
\int_{ B\left[ E(M)\right]  }d {\cal N}  \left( 
\xi \right) \
\varrho \left( \xi \right) \left( q \right)  .
\end{equation}
\end{lem}

\par\noindent${\it Proof.}\ \ $
{\it
The direct calculation
based on separation
(\ref{cl-separation})
shows 
\begin{eqnarray}\nonumber
\bar \mu \left( p^* F\right) &=&
\int_{
\Gamma \left[ E(M)\right] 
}d {\cal M}  \left( \eta  \right)  \
\tilde \mu  \left( \eta  \right)  \left( p^*F  \left( \eta   \right) 
  \right)  \\ \nonumber
&=& \int_{\Gamma \left[ E(M)\right] }d {\cal M} \left( \eta [k] \right)  \
\int_M dv(x) \
\varrho \left( \xi \right)  \left( x, p  \left( \eta [k] \right)  (x)\right)  
F^{T^*M} \left( x,  p  \left( \eta  [k]\right)  (x) \right)   \\ \nonumber
&=& 
\int_{{\bf R}^N} d^Nk
\int_{B \left[ E(M)\right] }
d {\cal N}  \left( 
\xi \right)
\int_M dv(x) \
\sigma   [ k](x) \\ \nonumber
& & \ \ \ \ \ \ \ \ \  \ \ \ \ \ \ \ \ \  \times \
\varrho \left( \xi \right)  \left( x, p \left( 
\eta [k] \right) (x) \right)  
F^{T^*M} \left( x,  p\left( 
\eta [k] \right) (x) \right)  \\ \nonumber
&=& 
\int_{{\bf R}^N} d^Nk
\int_{B \left[ E(M)\right] }
d {\cal N}   \left( \xi\right) 
\int_M dv(x) \
\sigma   [ k](x) \\
& & \ \ \ \ \ \ \ \ \  \ \ \ \ \ \ \ \ \  \times \
\varrho \left( \xi \right) \left( x,\bar  p  [k]   (x) 
+p\left( \xi\right) (x) \right)  
F^{T^*M} \left( x,  \bar  p  [k]   (x) 
+p\left( \xi\right) (x) \right)  \\ \nonumber
&=& 
\int_{B \left[ E(M)\right]  }
d {\cal N}  \left( \xi \right)  \sum_{\alpha \in \Lambda_M }
\int_{\phi_{U_{\alpha }}\left( A_{\alpha }\right) }
d^Nk\wedge d^Nx \
\phi_{U_{\alpha }}^*\varrho \left( \xi \right) \left( x , k \right)  
\phi_{U_{\alpha }}^*F^{T^*M} \left( x,  k\right) 
\\
&=& 
\int_{B \left[ E(M)\right]  }
d {\cal N}  \left( \xi \right) 
\int_{T^*M} \omega^N (q) \
\varrho \left( \xi \right) \left( q\right)  
F^{T^*M} \left( q\right) ,
\end{eqnarray}
where 
$T^*M= \bigcup_{\alpha \in \Lambda_M }
A_{\alpha }$
is  the  disjoint union 
of $A_{\alpha } \in {\cal B}\left( {\cal O}_{T^*M} \right) $
such that
(1) $\pi \left( A_{\alpha }
\right)  \subset U_{\alpha }$ and that (2)
 $A_{\alpha }\cap A_{\beta }
= \emptyset $ for $\alpha \neq \beta \in \Lambda_M $
(consult {\it APPENDIX}).

If defining the probability function
$\rho^{T^*M} : T^*M \to {\bf R}$ such that
\begin{equation}
\rho^{T^*M} \left( q \right)  =
\int_{B \left[ E(M)\right]  }d {\cal N}  \left( \xi \right) \
\varrho \left( \xi \right) \left( q \right)  ,
\end{equation}
we can obtain this lemma.
}\hspace{\fill} { \fbox {}}\\

\subsection{Description of Time-Development}

Let us consider the time-development
of the functional $\bar 
\mu_t : C^1 \left( \Gamma \left( M\right) ,C\left( M\right)
\right) \to {\bf R}$
for $
p_t^{\tau } \left( \eta [ k] \right) = 
\bar  p _t^{\tau } [k] 
  + p  \left( \xi \right) $.
For the Jacobian-determinant
$ \sigma_t^{\tau }  [ k]  = det 
\left( {{\partial \bar p_{ti} ^{\tau } [ k] }
\over {\partial k_j}}\right) $,
the following relation holds:
\begin{eqnarray}
 \bar \mu_t \left( p^* F_t\right) &=&\int_{T^*M} 
\omega^N (q) \
\rho_t^{T^*M} \left( q \right)  
F^{T^*M} \left( q \right) \\
&=& \int_{{\bf R}^N} d^Nk
\int_M dv(x) \
\bar  \rho_t^{\tau }   [ k] (x) 
F^{T^*M} \left( x, \bar   p_t^{\tau } [k] (x) \right) 
  ,
\end{eqnarray}
where
\begin{equation}\label{classical probability}
\bar  \rho_t^{\tau }   [ k](x)   \surd =
\sigma_t ^{\tau } [k](x) 
\rho^{T^*M} \left( x,  \bar  p_t^{\tau } [k] (x) \right)  .
\end{equation}
The Jacobian-determinant
$ \sigma_t^{\tau }  [ k] $ satisfies
the following relation:
\begin{equation}
 {{d {\cal M}\left( \eta_t^{\tau }(\eta )\right) }\over 
{d {\cal M}(\eta ) }} 
  = {{ \sigma_t^{\tau }  [ k] }\over {\sigma  [ k] }} .
\end{equation}

Thus,
we can define
the reduced emergence-momentum
$ \bar {\cal J}_t  
\in   \bar q\left(M\right) ^*= q\left(M\right)  ^*/
B \left[ E(M)\right]  $ as follows:
\begin{equation}
\bar  {\cal J}_t \left( \bar  \eta [k] \right)  
= \left(  d^Nk \  \wedge dv \
\bar   \rho_t ^{\tau } [k]  \otimes 
\bar  p_t^\tau [k]  , d^Nk\ \wedge 
 dv \ \bar   \rho_t ^{\tau }  [k] \right) ;
\end{equation}
and we can define the functional 
$\bar {\cal F}_t \in C^{\infty }\left( 
  \bar q\left(  M\right) ^*  , {\bf R} \right)  $
as
\begin{eqnarray}
\label{function}
\bar {\cal F}_t  \left(  
\bar {\cal J}_t  \right) 
&=& \bar \mu_t
  \left( p^* F_t   
  \right) \\
&=&
\int_{{\bf R}^N}d^Nk \
\int_M dv(x) \
\bar    \rho_t ^{\tau } [k]  (x) 
F_t^{T^*M} \left( x ,\bar   p_t^{\tau }[  k]  (x) \right) ,
\end{eqnarray}
which is independent of  labeling time $\tau $.

Then, the
operator $\hat F_t^{cl } ={{\partial \bar {\cal F}_t}
\over {\partial \bar {\cal J}}}  \left( \bar {\cal J}_t\right) $ satisfies
\begin{equation}
\hat F_t  ^{cl}
= \left(    {{\partial F_t^{T^*M}  }\over { \partial p \ \ }}
\left(x ,\bar   p_t^{\tau } [k]
(x ) \right) ,
-L^{F_t^{T^*M}} \left(  x , {{\partial F_t^{T^*M}  }\over {\partial p \ \ }} 
\left( x , \bar  p_t^{\tau } [k]
(x ) \right)
  \right) \right)  ,
\end{equation}
where
\begin{equation}\label{F-lagrange}
  L^{F_t^{T^*M}}  \left( { x,{{\partial F_t^{T^*M}  }\over {\partial p \ \ }} 
\left( x, p \right) } \right) =
p \cdot {{\partial F _t^{T^*M}  }\over {\partial p \ \ }}\left( x, p \right)  
-F_t^{T^*M}   \left( x, p  \right) 
\end{equation}
is the Lagrangian if function $F_t$ is Hamiltonian $H_t$.
Thus, the following null-lagrangian relation
 can be obtained:\footnote{
The Lagrangian corresponding to this Lie-Poisson system
is $
\langle \bar {\cal J}_t  , \hat H _t ^{cl} \rangle 
-{\cal H}_t  
 \left( \bar {\cal J}_t  \right) 
$, while
the usual Lagrangian is $L^{H_t^{T^*M}}$. }
\begin{equation}
 \bar {\cal F}_t  
 \left( \bar {\cal J}_t  \right) 
= 
\langle \bar {\cal J}_t  , \hat F _t^{cl}  \rangle  .
\end{equation}
Besides, 
the normalization condition 
becomes 
\begin{equation}\label{normalize}
\bar {\cal I}\left( \bar {\cal J} _t \right) =1 
\ \ \ \ \ for \ \ \ \ \ 
\bar {\cal I}\left(\bar  {\cal J} _t \right) =
\int_{{\bf R}^N}d^Nk  \
\int_M dv(x) \ \bar   \rho_t^{\tau } [k]( x) 
.
\end{equation}

\begin{tm}
For Hamiltonian operator $\hat H_t  =
{{\partial {\cal H}_t}\over {\partial \bar {\cal J}}} 
  \left(\bar {\cal J} _t \right)\in   \bar q\left(  M\right) $,
the equation of motion
becomes Lie-Poisson equation:
\begin{equation}\label{H[X]cl}
{{\partial \bar {\cal J}_t}\over {\partial t}} 
= ad^*_{\hat H_t^{cl}  }\bar {\cal J}_t   ,
\end{equation}
that is  calculated 
as follows:
\begin{equation}
\label{density's}
{{\partial  } \over {\partial t}} 
\bar   \rho _t^{\tau }[k](x) 
= -\surd ^{-1}\partial_j \left(
{{\partial H^{ T^*M }_t}\over {\partial p_j}} 
\left( x, \bar  p_t^{\tau } [k](x)\right)
\bar   \rho  _t^{\tau }[k](x)  \surd  \right) ,
\end{equation}
\begin{eqnarray}\nonumber
{{\partial  } \over {\partial t}}
\left( \bar  \rho  _t^{\tau }[k](x) 
\bar    p_{tk}^{\tau } [k](x) \right)
\nonumber
&=&
 - \surd^{-1} \partial_j \left(
{{\partial H^{ T^*M }_t}\over {\partial p_j}} \left( x, 
\bar  p_t^{\tau } [k](x)\right) 
 \bar  \rho  _t^{\tau }[k](x)  \bar  p_{tk}^{\tau } [k](x)
\surd  \right) \\
\nonumber
&  &
-  \bar  \rho  _t^{\tau }[k](x)  \bar  p_{tj}^{\tau } [k](x) 
\partial_k  
\left( {{\partial H^{ T^*M }_t}\over {\partial p_j}} 
\left( x, \bar  p_t^{\tau } [k](x)\right)  \right) \\
\label{current's}
&  &
+ \bar  \rho  _t^{\tau }[k](x)  
\partial_k  L^{H^{ T^*M }_t } \left( x, \bar  p_t^{\tau } [k](x)\right)  .
\end{eqnarray}
\end{tm}
\par\noindent${\it Proof.}\ \ $
{\it
The above equation can be obtained from
the integration of general equations
(\ref{density's-q}) and (\ref{current's-q})
on the space $\Gamma_{U 0}$;
thereby, it proves the
reduced equation from original Lie-Poisson equation
(\ref{Hq}).
}\hspace{\fill} { \fbox {}}\\

As a most important result,
the following theorem
shows that Lie-Poisson equation
(\ref{H[X]cl}), or the set of
equations (\ref{density's})
and (\ref{current's}),
actually
represents 
the classical Liouville equation.

\begin{tm}\label{canonical equivalent}
Lie-Poisson equation (\ref{H[X]cl}) 
is equivalent to
the classical Liouville equation
for the probability density
function (PDF) $\rho_t^{T^*M} \in C^{\infty }(T^*M,{\bf R})$
of a particle on cotangent space $T^*M$:
\begin{equation}\label{equation for rho}
{{\partial } \over {\partial t }}\rho^{T^*M}_t  =
\{ \rho^{T^*M}_t , H^{T^*M} \} ,
\end{equation}
where the Poisson bracket 
$\{ \ , \ \} $  is defined for
every $A$, $B\in C^{\infty }(M)$ as
\begin{equation}
\{ A , B \} = {{\partial A}\over {\partial p_j}}
{{\partial B}\over {\partial x^j}}
-{{\partial B}\over {\partial p_j}}{{\partial A}\over {\partial x^j}} .
\end{equation}
\end{tm}
\par\noindent${\it Proof.}\ \ $
{\it Classical equation (\ref{equation for rho})
 is equivalent to
 the
canonical equations of motion
through the local expression such that $\phi_{U_{\alpha }}
\left( q_t \right) = \left( x_t , p_t \right) $ for
the bundle mapping $\phi_{U_{\alpha }}: \pi^{-1}(U_{\alpha }) \to 
U_{\alpha } \times {\bf R}^N$:
\begin{equation}\label{classical canonical p:app}
{{dp_{jt}}\over {dt}}= 
-{{\partial  H^{T^*M} }
\over {\partial x^j}}(x_t ,p_t)
\ \ \ \ \ \ \ \ \ \ \ \ \ \ \ \ \ \ \ 
{{dx^j_t}\over {dt}} =
{{\partial H^{T^*M} }\over {\partial p_j}}(x_t ,p_t) .
\end{equation}

If 
$ q_t  
=  \left( x_t ,  \bar  p_{t} ^{\tau }[k]
(x_t ) \right)
$
satisfies
canonical equations  of motion
(\ref{classical canonical p:app}),
the above equation of motion
induces 
\begin{equation}\label{p's eq:app}
{{\partial\bar   p_{tk}^{\tau } } \over {\partial t}}[k](x)
= -{{\partial  H^{T^*M} }\over {\partial x^k }}
 \left( x  ,\bar   p_t^{\tau } [k] (x)
\right) 
-{{\partial H^{T^*M} }\over {\partial p_j}}
 \left( x  ,\bar   p_t^{\tau } [k] (x)
\right)
\partial_j  \bar  p_{tk}^{\tau } [k](x) ,
\end{equation}
then 
relation (\ref{classical probability})
satisfies the following equation:
\begin{eqnarray}\nonumber
{{\partial  } \over {\partial t}} \bar \rho^{\tau } _t[ k](x) 
&=&  \surd ^{-1}  
 \partial_j
\left(  \sigma_t^{\tau } [k] (x)
{{\partial H^{T^*M} }\over {\partial p_j}} \left( x,
\bar   p_t^{\tau } [k](x)\right) 
\right) \rho^{T^*M}_t (x ,\bar  p_t^{\tau }[k] (x) ) \\
\nonumber
&  &
+ \surd ^{-1}  
 \sigma_t^{\tau } [k] (x)
{{\partial   \rho^{T^*M}_t } \over {\partial t}} (x ,\bar  p_t^{\tau }[k] (x) )\\
\nonumber
&  &  
- \surd ^{-1}  \sigma_t^{\tau } [k] (x)
{{\partial H^{T^*M} }\over {\partial x^j}} 
\left(  x, \bar  p_t^{\tau } [ k](x)\right)
  {{\partial  \rho^{T^*M}_t  } \over {\partial p_j}}  (x ,
\bar  p_t^{\tau }[ k] (x) ) \\
\nonumber
&  &
- \surd ^{-1}  \sigma_t^{\tau } [ k] (x)
{{\partial H^{T^*M} }\over {\partial p_j}} \left( x, \bar  p_t^{\tau } [ k](x)\right)
\\
\nonumber
&  &
\times
 \partial_j \bar   p_{tk} [ k](x)
 {{\partial   \rho^{T^*M}_t } \over {\partial p_k}}  (x ,\bar  p_t^{\tau }[ k] (x) )\\
\label{density's eq:app}
&=& - \surd ^{-1} \partial_j   \left(
{{\partial H^{T^*M} }\over {\partial p_j}} \left( x, \bar  p_t^{\tau } [ k](x)\right)
 \rho^{\tau } _t[k](x)  \surd \right) .
\end{eqnarray}
Equations (\ref{p's eq:app}) and (\ref{density's eq:app})
lead to the following equation:
\begin{eqnarray}\nonumber
{{\partial  } \over {\partial t}}
\{ \bar \rho^{\tau } _t[k](x)  \bar  p_{tk}^{\tau }[k](x) \}
&=&   
 -\bar  p_{tk} [k](x)  \surd ^{-1} \partial_j \left(
{{\partial H^{T^*M} }\over {\partial p_j}} \left( x, \bar  p_t^{\tau } 
[k](x)\right)  
 \bar \rho^{\tau } _t[k](x) \surd \right) \\
\nonumber
&  &
 - \bar \rho _t^{\tau }[k](x) {{\partial 
 H^{T^*M} }\over {\partial x^k }}
\left( x, \bar  p_t^{\tau } [k](x) \right) \\
\nonumber
&  &
- \bar \rho^{\tau } _t[k](x) 
{{\partial H^{T^*M} }\over {\partial p_j}} \left( x, \bar  p_t^{\tau } [k](x)\right)
\partial_j  \bar  p_{tk}^{\tau } [k](x) \\
\nonumber
&=&
 -  \surd ^{-1} \partial_j \left(
{{\partial H^{T^*M} }\over {\partial p_j}} \left( x, 
\bar  p_t^{\tau } [k](x)\right)  
\bar  \rho^{\tau } _t[k](x) \bar  p_{tk}^{\tau } [k](x) \surd \right)\\
\nonumber
&  &
- \bar  \rho^{\tau } _t[k](x) 
\left\{ \bar   p_{tj}^{\tau } [k](x) 
\partial_k  
\left( {{\partial  H^{T^*M} }\over {\partial p_j}} 
\left( x, \bar  p_t^{\tau } [k](x)\right)  \right) \right.
\\
\label{current's eq:app}
&  & \left.
+ \partial_k  L^H  \left( x, \bar  p_t^{\tau } [k](x)\right) 
\right\}  .
\end{eqnarray}
Equations (\ref{density's eq:app}) and
(\ref{current's eq:app})
are equivalent to equations (\ref{density's}) and
(\ref{current's});
thereby, canonical equation (\ref{equation for rho})
is equivalent to
Lie-Poisson equation (\ref{H[X]cl}).
}\hspace{\fill} { \fbox {}}\\

The above discussion has
a special example
of the following Hamiltonian:
\begin{equation}\label{special Hamiltonian}
H^{ T^*M }_t   \left( x , p   \right)
= 
g^{ij}  (x)\left( p_{i}   
+ A_i  
\right)
\left( p_{j}   + A_j  
\right)
+U   (x) ,
\end{equation}
where corresponding Hamiltonian
operator $ \hat H_t  $
is calculated 
as 
\begin{equation}
\hat H_t [k]
= \left( 
 g^{ji} \left( \bar  p _{ti}[k]  + A_i   \right)
 \partial_j ,
 -  g^{ji}  \bar  p _{tj}[k] \bar   p _{ti} [k]   
+g^{ji} A_j  A_i   +U  
 \right)  ; 
\end{equation}
thereby, equation (\ref{H[X]cl})
is described for special Hamiltonian (\ref{special Hamiltonian})
 as
\begin{eqnarray}
\nonumber
{ {\partial } \over  {\partial t}  } 
\left( \bar  \rho _t^{\tau }[k] (x) \bar  p_{tj}^{\tau }[k](x) \right)  &=& -
 \surd ^{-1} \partial_i 
\left\{ g^{ik}(x)
\left( \bar  p_{tk}^{\tau }[k](x) 
 +A_k(x) \right)  \bar \rho _t^{\tau } [k](x)
\bar  p_{tj}^{\tau } [k] (x) \surd  \right\} \\
\nonumber
&  &- \bar  \rho _t^{\tau } [k](x) \left( \partial_j 
 g^{ik}(x)  \right) 
\bar  p_{ti}^{\tau }[k] (x)
\bar  p_{tk}^{\tau }[k] (x)\\ \nonumber
&  &-
\left( \partial_j  g^{ik}(x) A_k 
(x) \right) \bar  \rho _t^{\tau }[k] (x) \bar  p_{ti}^{\tau }[k] (x) \\ 
\label{motion equation 1}
&  &  -
\bar  \rho _t^{\tau }[k] (x) \partial_j 
\left\{  U(x) 
+ g^{ik}(x) A_i(x)  A_k (x) \right\} ,
\end{eqnarray}
\begin{equation}\label{motion equation 2}
{ {\partial } \over  {\partial t}  } \bar  \rho _t^{\tau }[k] (x)   = 
- \surd ^{-1} \partial_i 
\left\{ g^{ik}(x)\left( \bar  p_{tk}[k](x) 
+ A_k (x) \right) \bar  \rho _t^{\tau }[k] (x) \surd  \right\} .
\end{equation}

For
$  \bar {\cal U}_t  
 \in \bar Q \left(  M\right) $ such that
$
{{\partial  
\bar  {\cal U}_t  }\over {\partial t}}  
\circ  \bar  {\cal U}_t     ^{-1}= 
\hat H_t^{cl}  
 \in   \bar q\left(  M\right)  $,
let us introduce operators
\begin{eqnarray}
  \tilde H_t^{cl}   =& Ad^{-1}_{\bar {\cal U}_t   }
\hat H_t^{cl}  ,\\
 \tilde F_t ^{cl} =& Ad^{-1}_{\bar {\cal U}_t   }
\hat F_t^{cl} ,
\end{eqnarray}
which induces the following equation
equivalent to
equation (\ref{H[X]cl}):
\begin{equation}
{{\partial }\over {\partial t}}\tilde F_t^{cl}  = 
\left[ \tilde H _t ^{cl},\tilde F_t^{cl}  \right] 
+\widetilde{\left( {{\partial F_t    }
\over {\partial t}} \right) }^{cl} .
\end{equation}
This expression
of the equations of motion
coincides with
the following Poisson equation
because of
Theorem \ref{canonical equivalent}:
\begin{equation}\label{cl. Poisson eq.}
{{d  } \over {d  t }} F_t^{T^*M}
=\left\{ H_t^{T^*M}, F_t^{T^*M} \right\}     +
{{\partial F_t^{T^*M}} \over {\partial  t }}     .
\end{equation}

As discussed in Section 3,
 if a  group action of Lie group
$ Q(M)$
keeps the Hamiltonian $\bar {\cal H}_t: \bar q(M)^* \to {\bf R}$  
invariant,
there exists an invariant charge 
function
$Q ^{T^*M}\in C^{\infty }(T^*M)$ and the induced
function $\bar {\cal Q}: \bar q(M)^* \to {\bf R}$ such that
\begin{equation}\label{charge-invariant(cl)}
\left[ \hat H _t^{cl}, \hat Q^{cl} \right]  = 0 ,
\end{equation}
where $\hat Q^{cl}  $ is expressed as
\begin{equation}
\hat Q ^{cl}
= \left(  {{\partial  Q^{T^*M}}\over {\partial p}}
\left(  x, \bar p_t^{\tau }  [k](x) \right) ,
- p  (\eta )  \cdot    {{\partial  Q^{T^*M}}\over {\partial p}}
\left(  x, \bar p_t^{\tau }  [k](x) \right)
+ Q^{T^*M}   \left(  x, \bar p_t^{\tau }  [k](x) \right)
\right)  .
\end{equation}
Relation (\ref{charge-invariant(cl)})
is equivalent to the following 
convolution relation:
\begin{equation}
\left\{  H _t^{T^*M},   Q^{T^*M} \right\} = 0 .
\end{equation}

In the argument so far on the dynamical construction
of classical mechanics,
the introduced infinite-dimensional freedom
of the background $B\left[ E(M)\right] $
seems to be redundant, while they appear as a natural consequence
of the general theory on protomechanics
discussed in the previous section.
In fact, it is really true that
one can directly induce classical mechanics as the dynamics of the 
Lagrange foliations of $T^*M$ in $L \left( T^*M \right) $.
In the next section, however,
it is observed that we will encounter difficulties
without those freedom
if moving onto the dynamical construction
of quantum mechanics.

\section{DEDUCTION OF QUANTUM MECHANICS}

In canonical quantum mechanics,
the  state of a particle 
on  manifold $M$
can be
represented as a position  in
the Hilbert space ${\cal H}(M)$ 
of all the $L_2$-functions over $M$.
In this section,  we will reproduce
the quantum equation of motion
from the general theory
presented in Section 4.
Let us here concentrate ourselves
on the case where $M$ is $N$-dimensional
manifold for simplicity,
though the discussion
below is still valid if
substituting an appropriate Hilbert space
 when
$M$ is infinite-dimensional
ILH-manifold\cite{Omori}.

\subsection{Description of Statistical-State}

Now,
we must be concentrated
on the case
where the physical
 functional $F\in C^{\infty }\left( \Lambda^1(M) , C^{\infty }(M) \right) $
depends  on the derivatives 
of the 1-form $p\left( 
\eta \right)\in \Lambda^1 (M) $
induced from $\eta \in \Gamma \left[ E(M) \right] $, then it
has the following expression:
\begin{equation}\label{semi-local F}
p^* F  \left( \eta \right) (x) = 
F^{Q} \left( x,  p\left( \eta \right) (x) ,
D p\left( \eta \right) (x) ,
... , D^{n} p\left( \eta \right) (x) ,...\right)   .
\end{equation}

Let us assume that $M$ has a finite
covering $M= \bigcup_{\alpha \in \Lambda_M}U_{\alpha }$
for the mathematical simplicity
such that
$\Lambda_M = \left\{ 1,  2, ... , \Lambda \right\} $
for some $\Lambda \in {\bf R}$,
and choose a  
coordinate system  
$\left( U_{\alpha } , {\bf x}_{\alpha  } 
\right) _{\alpha \in \Lambda_M}$.
Let us further choose
a reference set $U \subset U_{\alpha }$
such that $v(U) \neq 0$
and consider the set $\Gamma^{\hbar }_{U k}\left[ E(M)\right] $ of  the
$C^{\infty }$ sections of $E(M)$
for $ k= (k_1 ,..., k_N) \in {\bf R}^N$
such that\footnote{
As in classical mechanics,
to substitute
$\Gamma^{\hbar }_{U k}\left[ E(M)\right] = \left\{ 
\eta \in \Gamma \left[ E(M)\right] \  \left\vert
\ \int_{  U}dv(x) \ p_j\left( x
\right)  = \hbar k_j v\left( U \right) \right.
\right\} $ for
definition (\ref{gamma})
also induces the similar discussion
below,
while there exist a variety of the classification
methods that produce the same result.
}
\begin{equation}\label{gamma}
\Gamma^{\hbar }_{U k}\left[ E(M)\right]  = 
\left\{ \eta \in \Gamma \left[ E(M)\right] \  \left\vert
\ \sup_{  U} p_j\left( \eta
\right)  (x) = \hbar k_j \right.
\right\} .
\end{equation}
As in classical mechanics,
 we will simply denote
$\Gamma^{\hbar }_{U k}\left[ E(M)\right] $
as $\Gamma^{\hbar }_{ k}\left[ E(M)\right] $,
since 
$\Gamma^{\hbar }_{U k}\left[ E(M)\right] $
can be identified with
$\Gamma^{\hbar }_{U^{\prime } k}\left[ E(M)\right] $
for every two reference sets $U$ and $U^{\prime }\subset M$.

For every $\bar  p 
= p\circ \bar  \eta  \in L\left( T^*M \right) $
such that 
$\bar  \eta [k] \in \Gamma^{\hbar }_{  k}\left[ E(M)\right]  $,
it is further possible to
separate an element $\eta [k] \in \Gamma^{\hbar }_{  k}\left[ E(M)\right] $
for a
$\xi\in  \Gamma^{\hbar }_{ 0}\left[ E(M)\right]  $ as
\begin{equation}
\eta [k]= \bar  \eta [k] \cdot \xi ,
\end{equation}
or to 
separate 
momentum
$p \left( \eta [ k] \right) $ as
\begin{equation}\label{q-separation}
p \left( \eta [ k] \right) 
= \bar  p    [ k]  
  + p  \left( \xi \right) .
\end{equation}
The emergence density $ \rho    \left( \eta [k]\right) 
$ can have  
the same expression as the classical one (\ref{rho=rho})
for the  function $ \varrho  \left( \xi \right)   \in  C^{\infty }
\left( T^*M,{\bf R}\right) $ on $T^*M$
since $C_{q}\left( \Gamma \right) ^*
\subset C_{cl}\left( \Gamma \right) ^*$:
\begin{equation}\label{rho=rho-q}
 \rho    \left( \eta [k]\right) 
(x) \surd 
=  
\varrho  \left( \xi \right)   \left( x, p \left( \eta [k]\right) (x) \right) ,
\end{equation}
which has only the restricted values
if
compared
with  the classical 
emergence  density; it
sometimes causes the discrete spectra of
the wave-function in canonical quantum mechanics.
We call the set
$ B^{\hbar }\left[ E(M)\right]  
= \Gamma^{\hbar }_{0}\left[ E(M)\right]  $ as
the {\bf back ground} of $L\left( T^*M \right) $
for quantum mechanics.
For the  measure
${\cal N}$ on $ B^{\hbar }\left[ E(M)\right]   $
for the $\sigma $-algebra induced from that of $\Gamma \left[ E(M)\right] $:
\begin{equation}
d {\cal M} \left( \eta [k]\right) \
d v (x)
= d^Nk d {\cal N}  \left(  \xi \right) 
 d v (x) \ \sigma [k](x)\  .
\end{equation}

 Let us next consider 
the disjoint union 
$M = \bigcup_{\alpha \in \Lambda_M }
A_{\alpha }$
for $A_{\alpha } \in {\cal B}\left( {\cal O}_{E(M)} \right) $
such that
(1) $\pi \left( A_{\alpha }
\right)  \subset U_{\alpha }$ and that (2)
 $A_{\alpha }\cap A_{\beta }
= \emptyset $ for $\alpha \neq \beta \in \Lambda_M $
(consult {\it APPENDIX}).
Thus,
every  section $
\eta \in \Gamma \left[ E(M)\right] $ has 
some $k\in {\bf R}^N$
such that $\eta =\eta [k]\in\Gamma^{\hbar }_{ k}\left[ E(M)\right] $; and,
it will be separated
into the product
 of a   $\xi \in B^{\hbar }\left[ E(M)\right] $
and the fixed $\bar \eta [k] =  
e^{2i \{ k_{  j} x^j 
+  \zeta   \} }\in \Gamma_{ k}\left[ E(M)\right] $
 that induces one of the Lagrange foliation $ \bar p
= p \circ \bar \eta \in L\left( T^*M \right) $:
\begin{eqnarray}
\eta \left[ k\right] &=& \sum_{\alpha \in A_{\alpha }}
 \chi_{A_{\alpha }}  \cdot
e^{2i \{ k_{  j} x^j 
+  \zeta   \} } \cdot \xi \\
&=&
\prod_{\alpha \in A_{\alpha }}
\left( 
e^{2i \{ k_{ j} x^j 
+  \zeta  \} } \cdot \xi \right) ^ {\chi_{A_{\alpha }}}
,
\end{eqnarray}
where the test function $\chi_{A_{\alpha }}:M \to {\bf R} $
satisfies 
\begin{equation}
\chi_{A_{\alpha }}(x)
= \left\{ 
{\matrix{1\cr
0\cr }\    \ } \right.
\matrix{{at\  x\in A_{\alpha }}\cr
{at\  x\notin A_{\alpha } }\cr
}  
\end{equation}
and has the projection property 
$ \chi_{A_{\alpha }}^2 = \chi_{A_{\alpha }}$.

If defining
the {\it window mapping} $\chi_{A_{\alpha }}^*:
C^{\infty }(M) \to L^1\left( {\bf R}^N\right) $
for any $f\in C^{\infty }(M)$
such that
\begin{equation}
\chi_{A_{\alpha }}^* f \left( {\bf x}\right) =
\left\{ 
{\matrix{\varphi_{\alpha }^* f \left(
 {\bf x}\right)\cr
0\cr }\    \ } \right.
\matrix{{at\  {\bf x}\in \varphi_{\alpha }\left( A_{\alpha } \right) }\cr
{at\   {\bf x}\notin \varphi_{\alpha }\left( A_{\alpha } \right)  }\cr
}  ,
\end{equation}
we can {\it locally} transform  the function $\rho [k]  \left(
\xi \right)
= \sigma [k] \rho \left( 
\eta \left[  k\right]  \right)   \surd  $
into Fourier coefficients as follows:
\begin{equation}
\chi_{A_{\alpha }}^*
\rho [k] \left(
\xi \right)  \  \left( {\bf x}\right) 
 = 
\int_{{\bf R}^{N }}  d^{N}k^{\prime } \
 \tilde \varrho_{\alpha } \left( \xi \right)
  \left(  {{2k
+ k^{\prime }}\over 2},  {{2k- 
k^{\prime } }\over 2} \right)
e^{ik^{\prime }  {\bf x} ^j } ,
\end{equation}
where introduced function
 $\tilde \varrho_\alpha  $
should satisfies
\begin{equation}
\tilde \varrho_\alpha \left( \xi \right) 
 (k, k^{\prime }) ^* =\tilde \varrho_\alpha 
\left( \xi \right)  ( k^{\prime },k),
\end{equation}
for the value $\rho [k] \left(
\xi \right) (x) $ is real at every $x\in M$;
thereby,
the collective expression
gives
\begin{eqnarray}
\rho   \left[  k\right]   \left(
\xi \right)
 &= & \sum_{\alpha \in A_{\alpha }}
 \chi_{A_{\alpha }}  \cdot
\int_{{\bf R}^{N }}  d^{N}k^{\prime } \
 \tilde \varrho  _\alpha \left( \xi \right)
  \left(  {{2k
+ k^{\prime }}\over 2},  {{2k- 
k^{\prime } }\over 2} \right)
e^{ik^{\prime }x^j } \\
&= &
\int_{
{\bf R}^{N  }} 
 d^{N  } k^{\prime } \
  \tilde \varrho  
\left( \xi \right)  \left(  {{2 k
+  k^{\prime }}\over 2},  {{2 k- 
 k^{\prime }}\over 2} \right) 
\cdot  \eta  \left[ 
 k -{{  k^{\prime }}\over 2}
\right] ^{-{1\over 2}}   \eta      \left[ 
 k+ {{  k^{\prime }}\over 2} 
\right]  ^{ {1\over 2}}  ,
\end{eqnarray}
where
\begin{equation}
  \tilde \varrho  \left( \xi \right)
  \left(  {{2 k
+  k^{\prime }}\over 2},  {{2 k- 
 k^{\prime }}\over 2} \right) 
= \prod_{\alpha \in A_{\alpha }}
\left( 
 \tilde \varrho  _\alpha \left( \xi \right)
  \left(  {{2k
+ k^{\prime }}\over 2},  {{2k- 
k^{\prime }}\over 2} \right) \right)^{\chi_{A_{\alpha }} } .
\end{equation}

Let us introduce
the ketvector $\left\vert   k     \right\rangle $
and bravector $\left\langle   k     \right\vert $
such that
\begin{equation}
  \left\vert   k     \right\rangle 
= \prod_{\alpha \in \Lambda _M }
\left\vert   k   ,\alpha
 \right\rangle    \ \  , \ \ \ \ 
\left\langle   k     \right\vert 
= \prod_{\alpha \in \Lambda _M }
\left\langle   k  ,\alpha   \right\vert ,
\end{equation}
where the local vectors $\left\vert   k   ,\alpha
 \right\rangle $ and $\left\langle   k  ,\alpha   \right\vert $
satisfy
\begin{equation}
\left\langle    x \left\vert  
 k   ,\alpha    \right. \right\rangle 
=
e^{2i \{ k_{ j} x^j 
+  \zeta  \} \chi_{A_\alpha } }  \surd ^{-{1\over 2}} \ \ , \ \ \
\ \left\langle   k, \alpha  \left\vert  
x    \right. \right\rangle 
=
e^{2i \{ -k_{ j} x^j 
+  \zeta  \} \chi_{A_\alpha } } \surd ^{-{1\over 2}}.
\end{equation}
We can define 
the Hilbert space ${\cal H} \left( M\right) $ of  all the vectors
that can be expressed as a linear combination
of vectors $\{ \vert k \rangle \}_{k\in {\bf R}} $.
Now,
let us construct the {\it density matrix}
in the following definition.
\begin{de}\label{q-density}
The {\bf density matrix}  $\hat \rho  $
is an operator
such that
\begin{eqnarray}
\hat \rho 
&=&
\int_{B^{\hbar }\left[ E(M) \right]  }
d{\cal N} (\xi ) 
\int_{
{\bf R}^{N }} d^N  n
\int_{
{\bf R}^{N }} 
d^N n^{\prime } \
\tilde \varrho (\xi )  \left( n,  n^{\prime }  \right) 
\ \xi^{{1\over 2} } 
 \left\vert  n \right\rangle  
\left\langle  n^{\prime } \right\vert    \xi^{-{1\over 2} } \\
&=& \int_{B^{\hbar }\left[ E(M) \right]   }
d{\cal N} (\xi )  \int_{
{\bf R}^{N }} d^N  k \ \hat  \rho [k]  \left( \xi \right) ,
\end{eqnarray}
where
\begin{equation}
\hat \rho  \left[  k\right]  \left( \xi \right) = 
\int_{
{\bf R}^{N }} d^N k^{\prime } \
\tilde \varrho (\xi )  \left( k+  {k^{\prime }\over 2} ,
  k-  {k^{\prime }\over 2}  \right) 
\ \xi^{{1\over 2} } 
 \left\vert   k+  {k^{\prime }\over 2} \right\rangle  
\left\langle   k-  {k^{\prime }\over 2} \right\vert \xi^{-{1\over 2} } .
\end{equation}
\end{de}

Let ${\cal O} \left( M\right) $
be the set of all the hermite operators acting on
Hilbert space ${\cal H} \left( M\right) $,
which has
the bracket $\langle \ \ \rangle : {\cal O} \left( M\right)
\to {\bf R}$ for every hermite operator
$\hat {\bf F}  $ such that
\begin{equation}
\left\langle \hat  {\bf F}    \right\rangle
= \int_{{\bf R}^N}d^Nk \int_M dv(x)\ 
\left\langle  x \left\vert \hat {\bf F}      \right\vert x \right\rangle .
\end{equation}
Set ${\cal O} \left( M\right) $
becomes the algebra
with the  product, scalar product  and addition;
thereby,
we can consider
 the commutation 
and the anticommutaion
between   operators $\hat {\bf A}$, $\hat {\bf B} \in {\cal O} \left( M\right) $:
\begin{equation}
\left[ \hat {\bf A}, \hat {\bf B}\right]_{\pm }
=  \hat  {\bf A}    \hat  {\bf B}   
\pm  \hat  {\bf B}    \hat  {\bf A}  .
\end{equation}
Consider the 
momentum operator $\hat {\bf p} $
that satisfies the following relation
for any $ \left\vert \psi  \right\rangle
\in {\cal H}\left( M \right) $:
\begin{equation}
 \left\langle x \left\vert  \hat {\bf p}  
\right\vert \psi \right\rangle
=-i D
\left\langle x \left\vert \psi \right. \right\rangle ,
\end{equation}
where $D = \hbar dx^j \partial_j$ is the derivative
operator (\ref{D-der}).
Further,
the function operator $ \hat {\bf f}   $
induced from the function $f \in C^{\infty }(M)$
is an operator that satisfies the following relation
for any $ \left\vert \psi  \right\rangle
\in {\cal H}\left( M \right) $:
\begin{equation}
\left\langle x \left\vert  \hat {\bf f} 
\right\vert \psi \right\rangle
=  f(x) 
\left\langle x \left\vert \psi \right. \right\rangle .
\end{equation}
The following commutation relation
holds:
\begin{equation}\label{q-commute}
\left[  \hat {\bf p}_j , \hat {\bf f}   \right] _-
= {\hbar \over i } \widehat{ \partial_j{\bf f}} .
\end{equation}
Those operators $\hat {\bf f} $ and $\hat {\bf p} $
induces a variety of operators
in the form of their polynomials.
\begin{de}
The hermite operator $\hat  {\bf F} $ is
called  an
{\bf observable},
if it 
can be represented as the polynomial
of the momentum operators $\hat p $
weighted
with  function operators $\hat {\bf f}_n^j $
independent of $k$ such that
\begin{equation}
\hat {\bf F}  =  \sum_{n=0}^{\infty } 
\left[ \hat {\bf f}_n^j  ,  \hat {\bf p}^n_{j} \right]_+ .
\end{equation}
\end{de}
The following lemma shows
that every {\it observable} has its own
physical functional.

\begin{lem}\label{q-gen}
Every observable
$\hat  {\bf F}  $ has 
a corresponding functional $F : \Gamma [E] \to C^{\infty }(M)$:
\begin{equation}
\bar \mu \left( p^*F  \right)  =
\left\langle \hat\rho  \ \hat {\bf F} 
\right\rangle .
\end{equation}
\end{lem}
\par\noindent${\it Proof.}\ \ $ 
{\it 
There are corresponding
functionals $g_{nl}^j: \Lambda^1(M) \to C(M)$
($l\in \{ 1, 2, ..., n \} $)
such that
\begin{eqnarray}\nonumber
\left\langle \hat\rho \ \left[ \hat {\bf f}_n^j  ,  \hat {\bf p}^n_{j} \right]_+
\right\rangle 
&=&
\int_{B^{\hbar }\left[ E(M) \right]   }
d{\cal N} (\xi ) 
\int_{{\bf R}^{N }} d^N  n
\int_{{\bf R}^{N }} d^N n^{\prime } \
\tilde  \varrho (\xi )  \left( n,  n^{\prime }  \right) 
\left\langle  n^{\prime } \left\vert   \ \xi^{-{1\over 2}}
\ \left[ \hat {\bf f}_n^j  ,  \hat {\bf p}^n_{j} \right]_+ \
\xi^{{1\over 2}} 
 \right\vert  n \right\rangle \\ \nonumber
&=&
\int_{B^{\hbar }\left[ E(M) \right]  }
d{\cal N} (\xi ) 
\int_{ {\bf R}^{N }}d^N  k
\int_{
{\bf R}^{N }} 
d^N k^{\prime } 
\\ \nonumber
&  & \times
\sum_{\alpha \in \Lambda_M}\int_{U_{\alpha }}  d^Nx \ 
\tilde  \varrho  
 (\xi ) \left( k-{k^{\prime }\over 2}, k+{k^{\prime }\over 2}
\right) e^{ik^{\prime }_jx^j }
  \left\{ 
\sum_{l= 0 }^{n}
g_{nl}^j\left( p \left( \eta [k] \right) \right) (x) k_j^{\prime l}\right\}
\\ \nonumber
&=&
 \int_{B^{\hbar }\left[ E(M) \right] }
d{\cal N} (\xi ) 
\int_{
 {\bf R}^{N }   } d^N  k
\int_{
{\bf R}^{N }} d^N k^{\prime } 
\\ \nonumber
&  & \times
\sum_{\alpha \in \Lambda_M}\int_{U_{\alpha }}  d^Nx \   
\tilde  \varrho   (\xi ) 
  \left( k-{k^{\prime }\over 2}, k+{k^{\prime }\over 2}
\right) 
e^{ik^{\prime }_jx^j } 
 \left\{ 
\sum_{l= 0 }^{n} 
\left( -\hbar {{ \partial }\over {\partial x^j}}\right) ^l
g_{nl}^j\left( p \left( \eta [k] \right)  \right) (x)
\right\}  \\ \nonumber
&=&
 \int_{B^{\hbar }\left[ E(M) \right]  }
d{\cal N}(\xi ) 
\int_{
{\bf R}^{N }} d^N  k
\int_{M}  dv(x)  \ 
\rho \left(  \eta [k]\right) (x) \
p^* F_j^n  \left(  \eta [k]\right) 
 (x)\\ \nonumber
&=&
 \int_{\Gamma \left[ E(M)\right] }
d{\cal M}(\eta ) 
\int_{M }   dv (x) \ 
\rho (\eta ) (x) \  p^* F_j^n (\eta )  (x) \\ 
&=& \bar \mu \left( p^* F_j^n \right)
.\end{eqnarray}
where 
\begin{equation}
p^* F_j^n 
\left(  \eta  [k]
  \right)  (x)  =
\sum_{l= 0 }^{n} \left\{
\left( -\hbar {{ \partial }\over {\partial x^j}}\right) ^l
g_{nl}^j\left(  p\left(  \eta [k]\right)  \right) (x)\right\} .
\end{equation}
}\hspace{\fill} { \fbox {}}\\

\subsection{Description of Time-Development}

Now, 
we can describe a 
$     \eta_t^{\tau } \left( \eta [k] \right)
\in  \Gamma _{Uk}\left[ E(M)\right]  $
as 
\begin{eqnarray}
   \eta_t^{\tau } \left( \eta [k] \right)
&=&  
\sum_{\alpha \in A_{\alpha }}
 \chi_{A_{\alpha }}  \cdot
e^{2i \{ k_{\alpha j} x^j 
+  \zeta_t^{\tau }[k]  \} } \cdot \xi \\
&=&
\prod_{\alpha \in A_{\alpha }}
\left( 
e^{2i \{ k_{\alpha j} x^j 
+  \zeta _t^{\tau } [k]\} } \cdot \xi \right) ^ {\chi_{A_{\alpha }}}
 ,
\end{eqnarray}
where the
function
$ \zeta  _t ^{\tau } 
[  k  ] 
\in C^{\infty }\left( M \right) $
labeled by {\it labeling time}
$\tau  \leq t \in {\bf R}$ satisfies
\begin{equation}
 \zeta _{\tau }^{\tau }  [  k  ] =  \zeta \ \ \ \ : independent\ of\ k   ;
\end{equation}
thereby,
the momentum $  p_t^{\tau }\left( 
\eta   [  k]\right) =
\bar p_t^{\tau } [  k ] + p \left( 
\xi \right) 
\in \Lambda ^1 (M) $ for $\bar p^{\tau }_{t} = 
p^{\tau }_{t} \circ \bar \eta \in L\left( T^*M \right) $
satisfies the  Einstein-de Broglie relation:\footnote{
Relation (\ref{improvement}) is 
the most crucial improvement from
the corresponding relation
in previous letter \cite{Ono}.}
\begin{equation}\label{improvement}
\bar   p_t^{\tau } [   k ] =
-i {\hbar \over 2} 
\bar   \eta_t^{\tau }   [  k ]  ^{-1} d 
\bar   \eta_t ^{\tau }  [   k ]  .
\end{equation}

The   
density operator $\hat \rho_t^{\tau } [k] \left( \xi \right) $
is introduced as
\begin{equation}\label{tau-dep density op}
 \hat \rho_t^{\tau } [k]  \left( \xi \right)  =
\int_{
{\bf R}^{N }} d^N k^{\prime } \
\tilde \varrho_t^{\tau } (\xi )  \left( k+  {k^{\prime }\over 2} ,
  k-  {k^{\prime }\over 2}  \right) 
\ \xi^{{1\over 2} } 
 \left\vert   k+  {k^{\prime }\over 2} \right\rangle  
\left\langle   k-  {k^{\prime }\over 2} \right\vert \xi^{-{1\over 2} } ,
\end{equation}
which satisfies the following lemma.
\begin{lem}
\begin{equation}\label{quantum-sigma-1}
\hat \rho_t          =   \int_{\Gamma _U}d{\cal N}(\xi )
 \int_{{\bf R}^N} d^Nk \
 U^{\tau }_t [k]   \hat \rho_t^{\tau }  [k] \left( \xi \right)
 U^{\tau }_t [k]^{-1}  ,
\end{equation}
where
\begin{equation}
{ U^{\tau }_t} \left[  k\right]  =   e^{i \{ \zeta_t^{\tau } 
\left[  k\right]  - \zeta  \} }   .
\end{equation}
\end{lem}
\par\noindent${\it Proof.}\ \ $
{\it 
The direct calculation shows for 
the observable $\hat {\bf F}_t $
corresponding to every functional $F$
\begin{eqnarray}\nonumber
\left\langle \hat\rho_t 
\ \hat {\bf F}_t  \right\rangle  &=&
\bar \mu_t
\left( p^*F_t    \right) \\ \nonumber
&=& \int_{\Gamma \left[ E(M)\right] }d{\cal M}(\eta ) \
\int_M dv \ \rho_t^{\tau }
\left(  \eta \right)  
(x) \ p^*F_t
\left(  \eta _t^{\tau } \left(\eta \right)  
  \right)   \\ \nonumber
&=&  \int_{B\left[ E(M)\right] }
d{\cal N}\left( \xi \right)  
\int_{ {\bf R}^{N }} d^N  k 
\int_{M}  dv(x)  \ 
\rho_t^{\tau } [k] \left( \xi \right) (x)  \ p^*F
\left(  \eta_t^{\tau }  [k]
  \right) (x)
  \\ \nonumber
&=&
\int_{B \left[ E(M)\right] }
d{\cal N}\left( \xi \right) 
\int_{
{\bf R}^{N }} d^N  k
\int_{M}  dv(x)  \ 
\rho_t^{\tau } [k] \left( \xi \right) (x) \ p^*F
\left( \eta [k] \cdot e^{i \{ \zeta_t^{\tau } 
\left[  k\right]  - \zeta  \} }   .
  \right) (x) 
 \\ \nonumber
&=&
\int_{B \left[ E(M)\right] }
d{\cal N}\left( \xi \right) 
\int_{
{\bf R}^{N }} d^N  k
\int_{M}  dv(x)  \ \left\langle x
\left\vert \
{1\over 2}\left[ \ U^{\tau }_t [k]  \hat
\rho_t^{\tau } [k] \left( \xi \right)   U^{\tau }_t [k]^{-1}  
\ , \ \hat {\bf F}_t \right]_+ \ \right\vert x
\right\rangle \ 
\\
&=&
  \ \left\langle 
\left\{
\int_{B\left[ E(M)\right] }
d{\cal N}\left( \xi \right) 
\int_{
{\bf R}^{N }} d^N  k \
  U^{\tau }_t [k]  \hat
\rho_t^{\tau } [k]\left( \xi \right)   U^{\tau }_t [k]^{-1} \right\} \ 
  \hat {\bf F}_t    
\right\rangle \  .
\end{eqnarray}
}\hspace{\fill} { \fbox {}}\\
Relation (\ref{quantum-sigma-1})
represents
relation (\ref{q-measure-rel}):
\begin{equation}
\tilde \mu_t   (\eta ) 
= {{d{\cal M} (  \eta  )  }
\over {d{\cal M}\left(  \eta_t^{\tau \ -1} (  \eta  ) \right)
}} \cdot \tilde \mu_t^{\tau }\left(
 \eta_t^{\tau \ -1}  \left(   \eta  \right) \right) .
\end{equation}

Emergence-momentum $ {\cal J}_t^{\tau }
={\cal J}\left( \eta _t^{\tau }  \right) \in q(M)^*$
has the following expression:
\begin{eqnarray}
  {\cal J}_t^{\tau } &=& d^Nk
d{\cal N}\left( \xi \right) dv \ \left( \rho_t^{\tau }
[k]\left( \xi \right)    p_t^{\tau }\left( \eta [k] \right) ,
\rho_t^{\tau }
[k]\left( \xi \right) \right) \\
\label{J}
&=& d {\cal N}\left( \xi\right)
 d^N k \wedge dv \
\left( \ {1\over 2}\left\langle 
x \left\vert \left[  \hat \rho _t^{\tau }  [k]  \left( \xi \right)
 ,\hat {\bf p}_t^{\tau } [k]  \right] _+
\right\vert
x \right\rangle \ ,\
 \left\langle 
x \left\vert   \hat \rho_t^{\tau }  [k]  \left( \xi \right)
\right\vert
x \right\rangle \ \right) ,
\end{eqnarray}
where the momentum operator $\hat {\bf p}_t^{\tau } [k]$
satisfies
\begin{equation}
\hat {\bf p}_t^{\tau } [k]  =  U_t^{\tau } [k]^{-1} \ \hat {\bf p}\  U_t^{\tau } [k]  .
\end{equation}
The following calculus
of the fourier basis
for $2k_j = n_j + m_j $ justifies expression (\ref{J}):
\begin{eqnarray}\nonumber
e^{-i\{
n_j {\bf x}^j + \zeta_t^{\tau }[k]   \} } d
e^{+i\{
m_j {\bf x}^j +  \zeta_t^{\tau }[k]    \} } -
e^{+i\{
m_j {\bf x}^j +  \zeta_t^{\tau }[k]      \} } d
e^{-i\{
n_j {\bf x}^j +  \zeta_t^{\tau }[k]    \} }
=\ \ \ \ &  &\\ \nonumber
e^{-i\{
n_j {\bf x}^j +  \zeta_t^{\tau }[k]    \} } d
\left\{  e^{+i\{
(m_j+n_j ) {\bf x}^j +  2\zeta_t^{\tau }[k]    \} } 
\cdot
e^{-i\{
n_j {\bf x}^j +  \zeta_t^{\tau }[k]   \} }
\right\} -
e^{+i\{
m_j {\bf x}^j + \zeta_t^{\tau }[k]     \} } d
e^{-i\{
n_j {\bf x}^j +  \zeta_t^{\tau }[k]     \} }
=\ \ \ \ &  &\\
 e^{-i (
n_j- m _j ) {\bf x}^j   } 
\cdot  e^{-i\{ (
n_j+m_j ) {\bf x}^j + 2  \zeta_t^{\tau }[k]     \} } 
d  e^{ +i\{ (
n_j+m_j ) {\bf x}^j + 2  \zeta_t^{\tau }[k]    \} } .&  &
\end{eqnarray}
For Hamiltonian operator $\hat  H_t ^{\tau } =
{{\partial {\cal H}_t}\over {\partial \bar {\cal J}}} 
  \left(\bar {\cal J} _t^{\tau } \right)\in   q\left( M\right) $,
the equation of motion
is the Lie-Poisson equation 
\begin{equation}\label{H[X]qq}
{{\partial  {\cal J} _t^{\tau }}\over {\partial t}} 
= ad^*_{\hat H_t  }  {\cal J} _t ^{\tau }  ,
\end{equation}
that is  calculated 
as follows:
\begin{equation}
\label{density's-qq}
{{\partial  } \over {\partial t}} 
 \rho _t^{\tau }[k]\left( \xi \right)  (x) 
= -\surd ^{-1}\partial_j \left(
{{\partial H^{ T^*M }_t}\over {\partial p_j}} 
\left( x,  p_t^{\tau }\left( \eta [k] \right) (x)\right)
\rho  _t^{\tau }[k]\left( \xi \right)  (x)  \surd   \right) ,
\end{equation}
\begin{eqnarray}\nonumber
{{\partial  } \over {\partial t}}
\left( \rho  _t^{\tau }[k]\left( \xi \right)  (x) 
 p_{tk}^{\tau }\left( \eta [k] \right) (x) \right)
\nonumber
&=&
 - \surd^{-1} \partial_j \left(
{{\partial H^{ T^*M }_t}\over {\partial p_j}} \left( x, 
p_t^{\tau } \left( \eta [k] \right) (x)\right) 
 \rho  _t^{\tau }[k]\left( \xi \right)  (x)  p_{tk}^{\tau }
 \left( \eta [k] \right) (x) \surd  
\right) \\
\nonumber
&  &
-  \rho  _t^{\tau }[k]\left( \xi \right)  (x)  
p_{tj}^{\tau } \left( \eta [k] \right) (x) 
\partial_k  
\left( {{\partial H^{ T^*M }_t}\over {\partial p_j}} 
\left( x, p_t^{\tau } \left( \eta [k] \right) (x)\right)  \right) \\
\label{current's-qq}
&  &
+ \rho  _t^{\tau }[k]\left( \xi \right)  (x)  
\partial_k  L^{H^{ T^*M }_t } \left( x, p_t^{\tau } 
\left( \eta [k] \right) (x)\right)  .
\end{eqnarray}
Notice that the above expression
is still valid even if
Hamiltonian $H_t^{T^*M}$
has 
the ambiguity of the operator ordering
such as that for the Einstein gravity.

To elucidate the relationship between the present
theory and canonical quantum mechanics,
we will concentrate on the case
of the canonical Hamiltonian
having the following form:
\begin{equation}\label{q-canonical Hamiltonian}
H^{ T^*M }_t   \left( x , p   \right)
= {1\over 2}
h^{ij} \left( p_{i}   
+ A_{ti}  
\right)
\left( p_{j}   + A_{tj}  
\right)
+U  _t (x) ,
\end{equation}
where $d h^{ij} =0$.
Notice that almost all the canonical 
quantum theory
including the standard model
of the quantum field theory,
that have empirically been 
well-established, 
really belong to this class of  Hamiltonian systems.
For Hamiltonian (\ref{q-canonical Hamiltonian}),
we will define the Hamiltonian
operator $\hat {\bf H}_t$ as
\begin{equation}
\hat {\bf H}_t=
 {1\over 2}
\left( \hat p_i + A _{ti}   \right)
h^{ij }  \left( \hat p_j +A _{tj} \right)  + U_t ,
\end{equation}
or
$
\langle x \vert \hat {\bf H}_t
\vert \psi \rangle
= {\cal H} _t\langle x \vert  
 \psi \rangle 
$
where
\begin{equation}
 {\cal H} _t=  {1\over 2}
\left( -i \hbar \partial_i + A _{ti} (x) \right)
h^{ij }  \left( -i  \hbar \partial_j +A _{tj} (x) \right)  + U_t (x) .
\end{equation}
\begin{lem}
Lie-Poisson equation  (\ref{H[X]qq})
for Hamiltonian (\ref{q-canonical Hamiltonian})
induces the following equation:
\begin{eqnarray}\label{q-liouville 1}
i \hbar {\partial 
\over {\partial t}} 
\left\langle x \left\vert
\hat \rho^{\tau } _t 
[k] \left( \xi \right) 
\right\vert x \right\rangle
&=&
- \left\langle x \left\vert
 \left[   \hat \rho^{\tau } _t [k] \left( \xi \right)  ,
\hat {\bf H}^{\tau } _t [k]   \right]_- 
\right\vert x \right\rangle
\\ \label{q-liouville 2}
i \hbar {\partial 
\over {\partial t}}
\left\langle x \left\vert
{1\over 2}
\left[ \hat \rho^{\tau } _t 
[k] \left( \xi \right) ,
 \hat {\bf p}^{\tau } _t [k]   \right] _+
\right\vert x \right\rangle
&=&
- \left\langle x \left\vert
\left[ \ {1\over 2}
\left[ \hat \rho_t ^{\tau }  [k] \left( \xi \right) 
 \ , \ \hat {\bf H}^{\tau } _t [k]  \right]_-  \ , \ \hat {\bf p}^{\tau } _t [k]   \right]_+
\right\vert x \right\rangle \ .
\end{eqnarray}
\end{lem}
\par\noindent${\it Proof.}\ \ $ 
{\it 
If we define the operators:
\begin{eqnarray}
 \hat {\bf H} _{(0)}  &=& \left. {{1 } \over {2  }}
h^{ij}  \hat {\bf p}^{\tau } _{t i}[k]  \hat {\bf p}^{\tau } _{tj} [k] \right/ i\hbar \\
\hat  {\bf H} _{(1)}     &=& \left.
{{1 } \over {2  }}\{  \hat  {\bf A} _i h^{ij}  
\hat {\bf p}^{\tau } _{tj} [k] +   \hat {\bf p}^{\tau } _{ti} [k] h^{ij}  \hat  {\bf A} _j \}\right/
i\hbar 
\\
\hat  {\bf H} _{(2)}   &=& \left. \left( \hat  U  +
{{1} \over {2 }}h^{ij} \hat  {\bf A} _i  \hat  {\bf A} _j  \right) \right/ i\hbar ,
\end{eqnarray}
then Hamiltonian operator $\hat {\bf H} _t $
can be represented as
\begin{eqnarray}
\left. \hat {\bf H} _t\right/  i\hbar  &=& \hat  {\bf H} _{(0)}   
 +\hat  {\bf H} _{(1)}   +\hat  {\bf H} _{(2)}   .
\end{eqnarray}

Thus, for 
density operator $\hat \rho^{\tau } _t 
[k] \left( \xi \right)   $ defined as equation
 (\ref{tau-dep density op}),
\begin{equation}
{{-1}\over {2  i\hbar  } }\left\langle x \left\vert
\left[ \ 
\left[ \hat \rho_t ^{\tau }  [k] \left( \xi \right) 
 \ , \ \hat {\bf H}^{\tau } _t [k]  \right]_-  \ , \ \hat {\bf p}^{\tau } _t [k]   \right]_+
\right\vert x \right\rangle \ 
= term_{(1)}  \left( \hat  {\bf H} _{(0)}   \right) +term_{(1)} \left(\hat  {\bf H} _{(1)}   \right) 
 +term_{(1)} \left(\hat  {\bf H} _{(2)}  \right) ,
\end{equation}
where
\begin{eqnarray}\nonumber
term_{(1)} \left( \hat  {\bf H} _{(0)}   \right) &=& 
{{-1}\over {2i\hbar }} \left\langle x \left\vert
\left[ \ {1\over 2}
\left[ \hat \rho_t ^{\tau }  [k] \left( \xi \right) 
 \ , \  \hat  {\bf H} _{(0)}   \right]_-  \ , \ \hat {\bf p}^{\tau } _t [k]   \right]_+
\right\vert x \right\rangle \ 
\\ \nonumber
term_{(1)} \left(\hat  {\bf H} _{(1)}   \right) &=&
{{-1}\over {2i\hbar }}\left\langle x \left\vert
\left[ \ {1\over 2}
\left[ \hat \rho_t ^{\tau }  [k] \left( \xi \right) 
 \ , \   \hat  {\bf H} _{(1)}   \right]_-  \ , \ \hat {\bf p}^{\tau } _t [k]   \right]_+
\right\vert x \right\rangle \ 
\\ \nonumber
term_{(1)} \left(\hat  {\bf H} _{(2)}   \right) &=& 
{{-1}\over {2i\hbar }}\left\langle x \left\vert
\left[ \ {1\over 2}
\left[ \hat \rho_t ^{\tau }  [k] \left( \xi \right) 
 \ , \  \hat  {\bf H} _{(2)}     \right]_-  \ , \ \hat {\bf p}^{\tau } _t [k]   \right]_+
\right\vert x \right\rangle \ .
\end{eqnarray}

First term 
results 
\begin{eqnarray}
term_{(1)} \left( \hat  {\bf H} _{(0)}  \right) 
 &=&
 -  \partial_j \left\{ h^{ij}  p_{ti}\left( \eta [k ] \right)   
\rho^{\tau } _t [k] \left( \xi \right) p_{tk}\left( \eta [k ] \right)  \right\}
 dx^k
\end{eqnarray}
from
 the following computations:
\begin{eqnarray}
\left\langle x \left\vert \
\hat {\bf p}^{\tau } _{tk} [k]  
\ \hat \rho^{\tau } _t [k] \left( \xi \right)  \
\hat  {\bf H} _{(0)}     \ \right\vert x \right\rangle
 &=& {1\over 2}\left. 
\int_{{\bf R}^N}d^Nk^{\prime } \
\tilde \rho^{\tau } _t  \left( \xi \right) \left(
k+{k^{\prime }\over 2} ,k-{k^{\prime }\over 2}
\right) e ^{ik^{\prime }\cdot x}
 \ \right\{ \\
&   &  \left( p_{tk}\left( \eta [k ] \right)+  \hbar
 {k_k^{\prime }\over 2} \right)
h^{ij}\left( p_{ti}\left( \eta [k ] \right)-  
\hbar {k_i^{\prime }\over 2} \right)
\left( p_{tj}\left( \eta [k ] \right)-   \hbar
{k_j^{\prime }\over 2} \right)\\
&  & \left. 
+i\hbar \left( p_{tk}\left( \eta [k ] \right)+ 
\hbar {k_k^{\prime }\over 2} \right) h^{ij}
\partial_j\left( p_{ti}\left( \eta [k ] \right)- 
\hbar {k_i^{\prime }\over 2} \right)
\right\}  ;
\end{eqnarray}
\begin{eqnarray}
\left\langle x \left\vert
\
\hat  {\bf H} _{(0)}        \
  \hat \rho^{\tau } _t [k] \left( \xi \right) 
 \
\hat {\bf p}^{\tau } _{tk} [k]  \ \right\vert x \right\rangle
 &=& {1\over 2}\left. 
\int_{{\bf R}^N}d^Nk^{\prime } \
\tilde \rho ^{\tau } _t \left( \xi \right) \left(
k+{k^{\prime }\over 2} ,k-{k^{\prime }\over 2}
\right) e ^{ik^{\prime }\cdot x}
 \ \right\{  \\
&   &  \left( p_{tk}\left( \eta [k ] \right)-  \hbar
{k_k^{\prime }\over 2} \right)
h^{ij}\left( p_{ti}\left( \eta [k ] \right)+ \hbar {k_i^{\prime }\over 2} \right)
\left( p_{tj}\left( \eta [k ] \right)+ \hbar {k_j^{\prime }\over 2} \right)\\
&  & \left. 
-i\hbar  \left( p_{tk}\left( \eta [k ] \right)- \hbar {k_k^{\prime }\over 2} \right) h^{ij}
\partial_j\left( p_{ti}\left( \eta [k ] \right)+ \hbar {k_i^{\prime }\over 2} \right)
\right\}  ;
\end{eqnarray}
\begin{eqnarray}
\left\langle x \left\vert\
\hat \rho^{\tau } _t [k] \left( \xi \right)  \
\hat  {\bf H} _{(0)}     \
\hat {\bf p}^{\tau } _{tk} [k] \ \right\vert x \right\rangle 
&=& {1\over 2}\left. 
\int_{{\bf R}^N}d^Nk^{\prime } \
\tilde \rho ^{\tau } _t \left( \xi \right) \left(
k+{k^{\prime }\over 2} ,k-{k^{\prime }\over 2}
\right) e ^{ik^{\prime }\cdot x}
 \ \right\{ \\
&   &   \left( p_{tk}\left( \eta [k ] \right)- \hbar {k_k^{\prime }\over 2} \right)
h^{ij}\left( p_{ti}\left( \eta [k ] \right)- \hbar {k_i^{\prime }\over 2} \right)
\left( p_{tj}\left( \eta [k ] \right)-  \hbar {k_j^{\prime }\over 2} \right)\\
&   & +i\hbar  
\left( p_{tk}\left( \eta [k ] \right)- \hbar  {k_k^{\prime }\over 2} \right)
h^{ij}\partial_i \left( p_{tj}\left( \eta [k ] \right)- \hbar {k_j^{\prime }\over 2} \right)\\
&   & -\hbar^2
h^{ij}\partial_k\partial_i \left( p_{tj}\left( \eta [k ] \right) - \hbar {k_j^{\prime }\over 2} \right)\\
&   &+i\hbar\left.
h^{ij}\partial_k
\left\{ \left( p_{ti}\left( \eta [k ] \right)- \hbar {k_i^{\prime }\over 2} \right)
\left( p_{tj}\left( \eta [k ] \right)-  \hbar {k_j^{\prime }\over 2} \right) \right\}
\right\}  ;
\end{eqnarray}
\begin{eqnarray}
\left\langle x \left\vert \
\hat {\bf p}^{\tau } _{tk} [k]  \ \hat  {\bf H} _{(0)}    
\ \hat \rho^{\tau } _t [k] \left( \xi \right)  \
 \right\vert x \right\rangle 
&=&{1\over 2}\left. 
\int_{{\bf R}^N}d^Nk^{\prime } \
\tilde \rho^{\tau } _t  \left( \xi \right) \left(
k+{k^{\prime }\over 2} ,k-{k^{\prime }\over 2}
\right) e ^{ik^{\prime }\cdot x}
 \ \right\{ \\
&   & +\left( p_{tk}\left( \eta [k ] \right)+\hbar {k_k^{\prime }\over 2} \right)
h^{ij}\left( p_{ti}\left( \eta [k ] \right) +\hbar {k_i^{\prime }\over 2} \right)
\left( p_{tj}\left( \eta [k ] \right)+  \hbar {k_j^{\prime }\over 2} \right)\\
&   & - i\hbar  
\left( p_{tk}\left( \eta [k ] \right)+  \hbar {k_k^{\prime }\over 2} \right)
h^{ij}\partial_i \left( p_{tj}\left( \eta [k ] \right)+ \hbar {k_j^{\prime }\over 2} \right)\\
&   & -\hbar^2
h^{ij}\partial_k\partial_i \left( p_{tj}\left( \eta [k ] \right) + \hbar {k_j^{\prime }\over 2} \right)\\
&   &
-i\hbar \left.
h^{ij}\partial_k
\left\{ \left( p_{ti}\left( \eta [k ] \right)+ \hbar {k_i^{\prime }\over 2} \right)
\left( p_{tj}\left( \eta [k ] \right)+ \hbar  {k_j^{\prime }\over 2} \right) \right\}
\right\} .
\end{eqnarray}

Further,
\begin{eqnarray}\nonumber
term_{(1)} \left(\hat  {\bf H} _{(1)}    \right)  
&=&- \left\{
 \partial_i \left(
h^{ij}    A _j  
\rho^{\tau } _t [k] \left( \xi \right)   p_{tk}\left( \eta [k ] \right)
\right) \right. \\ 
&  &\left. +\rho^{\tau } _t [k] \left( \xi \right) 
 \left( \partial_k h^{ij}  A_j  \right)
 p_{ti}\left( \eta [k ] \right)  \right\}  dx^k;
\\ 
term_{(1)} \left(\hat  {\bf H} _{(2)}    \right)  
&=&
-  \rho^{\tau } _t [k] \left( \xi \right) 
 \partial_k\left( U  +
{{1} \over {2}}h^{ij}  A _i A _j
\right)  dx^k  .
\end{eqnarray}

Thus, second equation
(\ref{q-liouville 2}) in this lemma
becomes 
\begin{eqnarray}
{\partial \over {\partial t}} 
\left\{ \rho^{\tau } _t [k] \left( \xi \right) 
 p_{tk}\left( \eta [k ] \right) \right\}
&=&
-  \partial_j \left\{ h^{ij}  \left( p_{ti}\left( \eta [k ] \right)  + A _j  \right) 
\rho^{\tau } _t [k] \left( \xi \right) p_{tk}\left( \eta [k ] \right)  \right\}\\ 
&  &  +\rho^{\tau } _t [k] \left( \xi \right) 
p_{tj}\left( \eta [k ] \right)   
 \left( \partial_k h^{ij}  A_i  \right)
\\ 
&   & -  \rho^{\tau } _t [k] \left( \xi \right) 
\partial_k \left( U  +
{{1} \over {2}}h^{ij}  A _i A _j
\right)   ,
\end{eqnarray}
which is equivalent to
equation
(\ref{current's-qq}) for
Hamiltonian (\ref{q-canonical Hamiltonian}).

On the other
hand,
\begin{equation}
 {{-1}\over {i\hbar }}
\left\langle x \left\vert
\  
\left[ \hat \rho_t ^{\tau }  [k] \left( \xi \right) 
 \ , \ \hat {\bf H}^{\tau } _t [k]  \right]_-  \  
\right\vert x \right\rangle \ 
= term_{(2)}  \left( \hat  {\bf H} _{(0)}   \right) +term_{(2)} \left(\hat  {\bf H} _{(1)}   \right) 
 +term_{(2)} \left(\hat  {\bf H} _{(2)}  \right) ,
\end{equation}
where
\begin{eqnarray}\nonumber
term_{(2)} \left( \hat  {\bf H} _{(0)}   \right) &=& 
{{-1}\over {i\hbar }}\left\langle x \left\vert
\
\left[ \hat \rho_t ^{\tau }  [k] \left( \xi \right) 
 \ , \  \hat  {\bf H} _{(0)}   \right]_-  \ 
\right\vert x \right\rangle \ 
\\ \nonumber
term_{(2)} \left(\hat  {\bf H} _{(1)}   \right) &=&
{{-1}\over {i\hbar }}\left\langle x \left\vert
\
\left[ \hat \rho_t ^{\tau }  [k] \left( \xi \right) 
 \ , \   \hat  {\bf H} _{(1)}   \right]_-  \ 
\right\vert x \right\rangle \ 
\\ \nonumber
term_{(2)} \left(\hat  {\bf H} _{(2)}   \right) &=& 
{{-1}\over {i\hbar }}\left\langle x \left\vert
\
\left[ \hat \rho_t ^{\tau }  [k] \left( \xi \right) 
 \ , \  \hat  {\bf H} _{(2)}     \right]_-  \ 
\right\vert x \right\rangle \ .
\end{eqnarray}

Each term can be calculated as follows:
\begin{eqnarray}
term_{(2)} \left( \hat  {\bf H} _{(0)}   \right) 
 &=& {{-1}\over {2i\hbar }}\left. 
\int_{{\bf R}^N}d^Nk^{\prime } \
\tilde \rho^{\tau }_t \left( \xi \right) \left(
k+{k^{\prime }\over 2} ,k-{k^{\prime }\over 2}
\right) e ^{ik^{\prime }\cdot x}
 \ \right\{ \\
&   &
h^{ij}\left( p_{ti}\left( \eta [k ] \right)- \hbar {k_i^{\prime }\over 2} \right)
\left( p_{tj}\left( \eta [k ] \right)- \hbar {k_j^{\prime }\over 2} \right)\\
&  &
+i\hbar  h^{ij}
\partial_j\left( p_{ti}\left( \eta [k ] \right)- \hbar  {k_i^{\prime }\over 2} \right)\\
&  &
-h^{ij}\left( p_{ti}\left( \eta [k ] \right)+ \hbar {k_i^{\prime }\over 2} \right)
\left( p_{tj}\left( \eta [k ] \right)+ \hbar  {k_j^{\prime }\over 2} \right)
\\
&  & \left.
-i\hbar  h^{ij}
\partial_j\left( p_{ti}\left( \eta [k ] \right)- \hbar  {k_i^{\prime }\over 2} \right)
\right\} \\
&=&  - \partial_j
\left(  \rho^{\tau } _t [k] \left( \xi \right) 
h^{ij}
  p_{ti}\left( \eta [k ] \right)  \right)
\\ \nonumber
term_{(2)} \left( \hat  {\bf H} _{(1)}   \right) 
&=&
 -  \partial_i h^{ij} 
\left( A _j 
\rho^{\tau } _t [k] \left( \xi \right) 
\right) ;\\
term_{(2)} \left( \hat  {\bf H} _{(2)}   \right) &=&0 .
\end{eqnarray} 

Thus, first equation
(\ref{q-liouville 2}) in this lemma
becomes
\begin{equation}
{\partial \over {\partial t}} \rho^{\tau } _t [k] \left( \xi \right) 
=
-  \partial_j \left\{ h^{ij}  \left( p_{ti}\left( \eta [k ] \right)   
+ A _j  \right)
\rho^{\tau } _t [k] \left( \xi \right)   \right\}   ,
\end{equation}
which is equivalent to
equation
(\ref{density's-qq}) for
Hamiltonian (\ref{q-canonical Hamiltonian}).

Therefore,
Lie-Poisson equation  (\ref{H[X]qq})
proved to be equivalent to the equation set
(\ref{q-liouville 1}) and (\ref{q-liouville 2}) in this lemma.
}\hspace{\fill} { \fbox {}}\\

The above lemma
leads us to one of the main theorem
in the present paper,
declaring
that
Lie-Poisson equation  (\ref{H[X]qq})
for Hamiltonian (\ref{q-canonical Hamiltonian})
is
equivalent to the quantum Liouville equation.
\begin{tm}
\label{main theorem}
Lie-Poisson equation  (\ref{H[X]qq})
for Hamiltonian (\ref{q-canonical Hamiltonian})
is equivalent to the following quantum Liouville equation:
\begin{eqnarray}
{\partial \over {\partial t}} \hat \rho  _t  
= \left[   \hat \rho  _t  ,
\hat {\bf H} \right]_- /(-i \hbar ) .
\end{eqnarray}
\end{tm}
\par\noindent${\it Proof.}\ \ $
{\it 
The following computation
proves this theorem
based on the previous lemma:
\begin{eqnarray}\nonumber
{\partial \over {\partial t}}
\left\langle \hat \rho_t \ \hat {\bf F}_t \right\rangle
&=&
{\partial \over {\partial t}}
\left\langle \hat \rho_t ^{\tau } \ \hat {\bf F}_t \right\rangle
\\ \nonumber
&=&
\int_{\Gamma _U }
d{\cal N}\left( \xi \right) 
\int_{{\bf R}^N} d^N k
\int_M dv(x) \times \\  \nonumber
&  &
\left\{ \left\langle  x \left\vert  
  \hat {\bf F}^{\tau } _t [k]    \
\hat \rho_t^{\tau } [k] \left( \xi \right)   \ \hat {\bf H} ^{\tau } _t [k] 
\right\vert x \right\rangle
- \left\langle  x \left\vert  \hat {\bf H}^{\tau } _t [k]  \
\hat \rho_t^{\tau } [k] \left( \xi \right)   
\  \hat {\bf F}^{\tau } _t [k] 
\right\vert x \right\rangle \right. \\ \nonumber
&  &  
+ 
\left\langle  x \left\vert
\ \hat \rho_t ^{\tau }  [k] \left( \xi \right) \ 
 \ \right\vert x \right\rangle 
\ { {\partial p_t^{\tau }[k] (x)}\over {\partial t}} 
\cdot
{\cal D} F _t\left( \eta_t^{\tau } [k]  \right)  (x) 
\\ \nonumber
&  &  
\left.
+
\left\langle  x \left\vert
\ \hat \rho_t ^{\tau }  [k]  
 \ \right\vert x \right\rangle 
p^*
{ {\partial F _t }\over {\partial t}} \left( \eta_t^{\tau } 
[k]  \right) (x) \right\} \\ \nonumber
&=&
\int_{\Gamma _U }
d{\cal N}\left( \xi \right) 
\int_{{\bf R}^N} d^N k
\int_M dv(x) \times \\ \nonumber
&  &
 \left\{
\left\langle  x \left\vert  \left[
\hat \rho_t^{\tau } [k] \left( \xi \right)
 ,  \hat {\bf H}^{\tau } _t [k] 
\right]_-
\right\vert x \right\rangle \
 p^* F_t \left( \eta_t^{\tau } [k] \right) (x) 
\right.\\ \nonumber
&  & 
+ 
\left( 
{\partial \over {\partial t}} 
\left\langle  x \left\vert \
{1\over 2}
\left[
\ \hat \rho_t ^{\tau }  [k] \left( \xi \right) \ 
,  \hat {\bf p}^{\tau } _t [k]  \right] _+ \ \right\vert x \right\rangle 
\right)
\cdot
{\cal D} F_t \left( \eta_t^{\tau } [k]  \right)  (x)
\\ \nonumber
&  & 
- \left\langle  x \left\vert \
{{\partial \hat \rho_t ^{\tau }  [k] \left( \xi \right) }
\over {\partial t}}\ 
 \right\vert x    \right\rangle 
\  p_t^{\tau } [k] (x) 
\cdot
{\cal D} F_t \left( \eta_t^{\tau } [k]  \right)   (x)
\\ \nonumber
&  & \left.
+
\left\langle  x \left\vert
\ \hat \rho_t ^{\tau }  [k]  
 \ \right\vert x \right\rangle 
p^*
{ {\partial F _t }\over {\partial t}} \left( \eta_t^{\tau } 
[k]  \right)
 (x) \right\}  \\ \nonumber
&=&
\int_{\Gamma _U }
d{\cal N}\left( \xi \right) 
\int_{{\bf R}^N} d^N k
\int_M dv(x) \times \\ \nonumber
&  & 
 \left\{
\left\langle  x \left\vert
\left[
\ {1\over 2}
\left[
\ \hat \rho_t ^{\tau }  [k] \left( \xi \right) \ 
,  \hat {\bf p}^{\tau } _t [k] \right] _+ \ ,\ \hat {\bf H}^{\tau } _t [k] 
\right]_- \right\vert x \right\rangle 
\cdot
{\cal D} F_t \left( \eta_t^{\tau } [k] \right)  (x)
\right. \\ \nonumber
&  &  
\left\langle  x \left\vert  \left[
\hat \rho_t^{\tau } [k] \left( \xi \right)
\  , \ \hat {\bf H}^{\tau } _t [k] 
\right]_-
\right\vert x \right\rangle \
\left\{
 p^* F_t \left( \eta_t^{\tau } [k] \right) (x) 
-   p_t^{\tau } [k] (x) 
\cdot
{\cal D} F_t \left( \eta_t^{\tau } [k]
\right)  (x) \right\}
\\ \nonumber
&  & 
\left.
+
\left\langle  x \left\vert
\ \hat \rho_t ^{\tau }  [k]  
 \ \right\vert x \right\rangle 
p^*
{ {\partial F _t }\over {\partial t}} \left( \eta_t^{\tau } 
[k]  \right) (x) \right\} 
\\ 
&=&
\left\langle ad^*_{\hat H_t^{\tau }}{\cal J}_t^{\tau }
, \hat F_t^{\tau }
\right\rangle
+\left\langle {\cal J}_t
, { {\partial \hat F _t }\over {\partial t}}
\right\rangle \
.
\end{eqnarray}
}\hspace{\fill} { \fbox {}}\\

Now, 
the density matrix $\hat \rho_t  $
becomes the summation of
the pure sates $ \left\vert \psi_t^{(l ; \pm )}\right\rangle
\left\langle  \psi_t^{(l ; \pm )} \right\vert   $
for the set $ 
\left\{ \left\vert \psi_t^{(l ; \pm )}
\right\rangle \right\}_{l \in {\bf R}^{N}} $
of the orthonormal wave vectors  
such that 
$\left\langle  \left. \psi_t^{(l^{\prime };s^{\prime })}
\right\vert  \psi_t^{(l; s)}
\right\rangle  
=\delta (l^{\prime }-l ) \delta_{s, s^{\prime }}$:
\begin{equation}
\hat \rho_t  = \int_{\Lambda } 
d P_{+} (l)  
\left\vert \psi_t^{(l ; + )}
 \right\rangle
\left\langle  \psi_t^{(l ; + )}
 \right\vert 
-
\int_{\Lambda } 
d P_{-} (l)  
\left\vert \psi_t^{(l ; - )}
 \right\rangle
\left\langle  \psi_t^{(l ; - )}
 \right\vert  ,
\end{equation}
where  $ P_{\pm }   $
is a corresponding probability
measure 
on the space $\Lambda $ of 
a spectrum
and the employed integral
is the Stieltjes integral \cite{Neumann}.
If the system is open and has the continuous spectrum,
then it admits 
$\Lambda $ be the continuous superselection rules (CSRs).
The induced wave function has the following expression
for a $L^2$-function $\psi_t^{(l ; \pm )}  =\left\langle  x  
\left\vert \psi_t^{(l ; - )} \right.
 \right\rangle \in L^2(M)$:
\begin{equation}
\chi_{\alpha }^*\psi_t^{(l ; \pm )} (x )=
\int_{{\bf R}^N}d^Nk \
\tilde  \psi_{\alpha \ t}^{(l ; \pm )} (k) e^{i\{k_j{\bf x}^j +
\zeta_t (x) \} } .
\end{equation}
The existence of the probability
measure $P_-$ would be corresponding to
the existence of the  antiparticle for the elementary
quantum mechanics.

For example,
the motion of the particle
on a N-dimensional rectangle box  $[0, \pi ]^N $
needs the following boundary condition
on the verge of the box:
\begin{quote}
if $x_j = 0$ or $ \pi $  for some $j \in \{ 1,..., N\} $,
then $ \left\langle x \left\vert
\hat \rho_t   \right\vert x \right\rangle = 0 $,
\end{quote}
Density matrix $\hat \rho_t $
is the summation
of  integer-labeled pure states:
\begin{equation}
\hat \rho_t =  
 \sum_{  ( n ,  n^{\prime } ) \in  {\bf Z}^{2N}  }
\ 
\tilde \rho^{t} _{ t} ( n^{\prime }, n )
\left\vert    n  ; t \right\rangle
\left\langle    n^{\prime } ;  t  \right\vert .
\end{equation}

Let us now concentrate on the case
where $\hat \rho_t $ is a pure state in the following form:
\begin{equation}
\hat \rho_t =  \left\vert \psi_t \right\rangle
\left\langle  \psi_t \right\vert  ;
\end{equation}
there exists a wave function $\psi_t  \in L^2 (M)$
\begin{equation}
\psi_t (x)  =
\int_{{\bf R}^N} d^Nk  \
\tilde \psi_t   (k) e^{i\{
k_j {\bf x}^j + \zeta_t (x)  \} }  ,
\end{equation}
where
\begin{equation}
\tilde \rho_t^t (k, k^{\prime }) 
=  
\tilde  \psi_t (k)^* \tilde  \psi_t ( k^{\prime })
.
\end{equation}
Theorem \ref{main theorem}
introduces the Schr\"odinger equation as the following collorary.

\begin{cl}
\label{main theorem for psi}
Lie-Poisson equation  (\ref{H[X]qq})
for Hamiltonian (\ref{q-canonical Hamiltonian})
becomes the following
Schr\"odinger equation:
\begin{equation}\label{Srodinger}
 i \hbar \partial_t \psi_t     
=   {\cal H}
\psi_t   ,
\end{equation}
where
\begin{equation}
 {\cal H} = {1\over {2m}}\surd ^{-1}   
\left( -i \hbar \partial_i +  A _{ti}(x)\right)
g^{ij }(x)\surd  \left( -i  \hbar \partial_j + A _{tj}(x)\right)  +U_t (x) .
\end{equation}
\end{cl}
Therefore,
the presented theory induces not only
canonical, nonrelativistic quantum mechanics
but also the canonical, relativistic or nonrelativistic quantum
field theory if proliferated for the grassmanian field variables.
In addition,
Section 7 will discuss how the present theory 
also justifies
the regularization procedure
in the appropriate renormalization.

On the other hand,
if introducing the unitary transformation $\hat U_{t} = e^{it\hat {\bf H}_t }$,
Theorem \ref{main theorem} 
obtains the Heisenberg equation
for Heisenberg's representations 
$ \tilde {\bf H} _t = \hat U_{t} \hat {\bf H} _t \hat U_{t} ^{-1}$
and $ \tilde {\bf F} _t = \hat U_{t} \hat {\bf F} _t \hat U_{t} ^{-1}$:
\begin{eqnarray}\label{Heisenberg}
 {\partial \over {\partial t}} \tilde {\bf F} _t
= \left[   \tilde {\bf H} _t  ,
\tilde {\bf F} _t\right]_-  / (-i \hbar )+\widetilde{\left( {{\partial {\bf F} _t    }
\over {\partial t}} \right) },
\end{eqnarray}
since $ \hat \rho  _t =\hat U_t^{-1} \hat \rho  _0 \hat U_t  $.

As discussed in Section 3,
 if a  group action of Lie group
$ Q(M)$
keeps the Hamiltonian ${\cal H}_t: q(M)^* \to {\bf R}$  
invariant,
there exists an invariant charge 
functional
$Q : \Gamma \left[ E(M)\right] \to
C(M) $ and the induced
function ${\cal Q}: q(M)^* \to {\bf R}$ such that
\begin{equation}\label{charge-invariant(q)}
\left[ \hat H _t, \hat Q \right]  = 0 ,
\end{equation}
where $\hat Q $ is expressed as
\begin{equation}
\hat Q 
= \left(  {\cal D}_{\rho (\eta )} Q
\left(  p  (\eta ) \right) ,
- p  (\eta )  \cdot     {\cal D}_{\rho (\eta )} Q
\left(  p  (\eta ) \right)
+ Q   \left( p (\eta )  \right)
\right)  .
\end{equation}
Suppose that functional $p^*Q:\Gamma \left[E(M)\right]  \to C(M)$
has the canonical form such that
\begin{equation}
Q ^{T^*M}(x,p)= A^{ij} p_i p_j +  B(x)_i p_j + C(x) ,
\end{equation}
then the corresponding generator
is equivalent to the observable:
\begin{equation}
\hat {\bf Q} = A^{ij} \hat {\bf p}_i  \hat {\bf p}_j +  
\hat {\bf B} _i  \hat {\bf p}_j +  \hat {\bf p}_j \hat {\bf B} _i 
+ \hat  {\bf C} .
\end{equation}
In this case, relation (\ref{charge-invariant(q)})
has the canonical expression:
\begin{equation}\label{charge-invariant(q) op}
\left[ \hat {\bf H} _t, \hat {\bf Q} \right]  = 0 .
\end{equation}
Those operators can have the eigen values at the same time.

As shown so far,
protomechanics successfully deduced quantum mechanics
for the canonical Hamiltonians
that have no problem in the operator ordering,
and proves still valid
for the noncanonical Hamiltonian
that have the ambiguity of the operator ordering
in the ordinary quantum mechanics.
In the latter case,
the  infinitesimal
generator $\hat {\bf F}_t^{tr}$ corresponding to $\hat F
\in q(M)$
is not always equal to
observable $\hat {\bf F}_t$:
\begin{equation}
\hat {\bf F}_t\neq \hat {\bf F}_t^{tr} .
\end{equation}
If one tries to quantize the Einstein gravity,
he or she can proliferate the present theory
in a direct way by utilizing
Lie-Poisson equation (\ref{H[X]qq}).
But,
some calculation method
should be developed for this purpose
elsewhere.

\subsection{Interpretation of Spin}

It has been known that
a half-spin in quantum mechanics
{\it does} have
a classical analogy as a
rigid rotor  in classical
mechanics \cite{Holland}.\footnote{ 
The ignorance on
this fact may have prevented
 quantum  mechanics from the realistic interpretation
in general.}
Such a model represents the motion
of a particle on the three-dimensional orthogonal group $SO( 3 )
$.
A spinor is corresponding to
an element of the 
Lie-algebra $so(3)$
of $SO( 3 )
$,
which is equivalent to 
a right-(or left-)invariant
vector field over $SO(3)$.
This section
reviews such an interpretation
of a spin
in terms of the Euler angles or the
coordinates over  a three-dimensional
special
orthogonal group $SO(3)$;
and thus, it 
proves that the present theory is
applicable for the description of a half-spin, too.

Now,
let us consider the
particle motion in a three-dimensional
Euclidean space ${\bf R}^3$
with the polar coordinates
${\bf x}=(r, \theta ,\phi ) \in [0 , +\infty ) \times
[0 , 2\pi ) \times (0, \pi )$.
Lie group $SO(3)
$ acts on ${\cal J}_t= \left(\rho _t^{\tau } p^{\tau }_t
 , \rho_t^{\tau }   \right) $ by the coadjoint
action, where
an infinitesimal
generator $ M = M^j  \hat L_j 
\in so(3)
\subset q\left(M \right) $  ($M_j \in {\bf R}$, $ j\in \{1,2,3\}$) 
has an corresponding operator $ \hat {\bf M}=
 M^j \hat {\bf L}_j   \in su(2,{\bf C }) $
that satisfies
\begin{equation}
\left\langle  ad^*_{\hat M} {\cal J}_t ,
 \hat F \right\rangle   =  -i\hbar ^{-1}
  \left\langle    \left[ \hat \rho_t ,  \hat {\bf M}\right] _{-}
\ \hat  {\bf F}  \right\rangle  
 .
\end{equation}
Infinitesimal generator
 $ \hat L_j  $
has the following expression:
\begin{eqnarray}
\hat L_1 & =&    -sin \phi {\partial \over {\partial \theta }}
-\cot\theta \cos\phi   {\partial \over {\partial \phi }}   \
, \\
\hat L_2 & =&   \cos \phi {\partial \over {\partial \theta }}
-\cot\theta \sin\phi   {\partial \over {\partial \phi }}  
\ ,\\
\hat L_3 & =&   {\partial \over {\partial \phi }}   \ ;
\end{eqnarray}
It 
has an corresponding operator $ \hat {\bf M}=
 M^j \hat {\bf L}_j   \in su(2,{\bf C }) $
acting on the Hilbert space ${\cal H}(S^2)$
of all the single- or double-valued $L_2$ functions over $S^2$:
\begin{eqnarray}
\left\langle \theta , \phi
\left\vert \hat {\bf L}_j \right\vert \psi \right\rangle 
=   {\hbar \over i}
\hat L_j   \
\left\langle \theta , \phi
\left\vert  \psi \right. \right\rangle  ,
\end{eqnarray}
where $\vert  \psi \rangle \in {\cal H}(S^2)$.
Notice that these operators are 
hermite or self-conjugate,
$
\hat {\bf L}_j^{\dag }   = 
\hat {\bf L}_j $,
and induces
the angular momentum or
the integer spin
of the particle:
\begin{eqnarray}
\left\vert   \psi_t   \right\rangle
&=&  
\sum_{m =  -l }^{  l}
\ 
c_m^l(t)
\left\vert    l,m   \right\rangle\\
&  &for  \ \ \ \ \ 
 \left\langle \theta , \phi \right. 
\left\vert l; m\right\rangle
= Y^m_l(\theta , \phi ) ,
\end{eqnarray}
where
\begin{equation}
\left.
 \hat {\bf L}\cdot  \hat {\bf L}
\left\vert    l,m   \right.\right\rangle
= \hbar^2 l(l+1) 
\left\vert l; m\right\rangle \ \ , \ \ \ \
\left.
 \hat {\bf L}_3
\left\vert    l,m   \right.\right\rangle
=   \hbar m 
\left\vert l; m\right\rangle .
\end{equation}

If the Hamiltonian for
the motion in the three-dimensional Euclid space
has the following form
in a central field of force, it is invariant under the rotation about z-axis:
\begin{equation}\label{spin-hamiltonian}
H \left( x, p \right) = p^2 +x\cdot \left( p\times B \right) + U(r) ,
\end{equation}
where $r=\sqrt{x^2+y^2+z^2}\neq 0$.
Since this Hamiltonian has the canonical form,
the corresponding infinitesimal generator is
equivalent to the following quantum observable \cite{Dirac}:
\begin{equation}
\hat {\bf H} = \hat {\bf P_r}^2 +
 {{\hat {\bf L}\cdot \hat {\bf L}  }\over {r^2}}+ 
{1\over 2}
\left\{ \hat {\bf L}\cdot B +B\cdot \hat {\bf L} \right\} + U(r),
\end{equation}
where
\begin{equation}
\left\langle \theta , \phi , r
\left\vert \hat {\bf P_r} \right\vert \psi \right\rangle 
= -{\hbar \over {ir}}{\partial \over {\partial r} } r\left\langle \theta ,
 \phi , r \left\vert \psi \right.\right\rangle .
\end{equation}

To realize
the representation for a half-spin,
let us consider the  Hilbert spaces ${\cal H}(SO(3))$
of all the single- or double-valued
$L_2$  functions over $S^2$
which can be reduced to ${\cal H}(S^2)$.
On the classical level,
an infinitesimal generator $  N= N^j S_j$
of  $SO(3)$
is equivalent to a
left-(or right-)invariant vector field:
\begin{eqnarray}
\hat S_1 & =&   \hat L_1 +
{\hbar \over i}\cdot {{\cos \phi }\over {\sin \theta }} 
{{\partial }\over {\partial \chi }} 
, \\
\hat S_2 & =&   \hat L_2 +
{\hbar \over i}\cdot {{\sin \phi }\over {\sin \theta }} 
{{\partial }\over {\partial \chi }} 
\ ,\\
\hat S_3 & =&  \hat L_3   \ .
\end{eqnarray}
Notice that infinitesimal generator $  N= N^j S_j$
is also an element of
 the semidirect product
 $SO(3)\times C^{\infty }(M)$
of $SO(3)$ with the space $ C^{\infty }(S^2)$
of all the $C^{\infty }$ functions over $S^2$ excepting poles
$\theta = 0 , \pi $.
The corresponding operators ${\bf S}_j$ in quantum mechanics
to generators $S_j$
become
\begin{eqnarray}
\left\langle \theta , \phi ,\chi
\left\vert \hat {\bf S}_1 \right\vert \psi \right\rangle 
& =&   \left\{ {\hbar \over i}\hat L_1 +
{\hbar \over i}\cdot {{\cos \phi }\over {\sin \theta }} 
{{\partial }\over {\partial \chi }}   \right\} 
\ \left\langle \theta , \phi ,\chi
\left\vert  \psi \right. \right\rangle
, \\
\left\langle \theta , \phi ,\chi
\left\vert \hat {\bf S}_2 \right\vert \psi \right\rangle  & =& 
 \left\{ {\hbar \over i}\hat L_2
+ {\hbar \over i}\cdot {{\sin \phi }\over {\sin \theta }} 
{{\partial }\over {\partial \chi }} 
\right\} 
\ \left\langle \theta , \phi ,\chi
\left\vert  \psi \right. \right\rangle ,\\
\left\langle \theta , \phi ,\chi
\left\vert \hat {\bf S}_3 \right\vert \psi \right\rangle  & =&
{\hbar \over i}\hat L_3  
\ \left\langle \theta , \phi ,\chi
\left\vert  \psi \right. \right\rangle .
\end{eqnarray}
which has the following reduced
expression:
\begin{eqnarray}
\left\langle \theta , \phi
\left\vert \hat {\bf S}_1 \right\vert \psi \right\rangle 
& =&   \left\{ {\hbar \over i}\hat L_1 +
{\hbar \over 2}\cdot {{\cos \phi }\over {\sin \theta }}   \right\} 
\ \left\langle \theta , \phi
\left\vert  \psi \right. \right\rangle
, \\
\left\langle \theta , \phi
\left\vert \hat {\bf S}_2 \right\vert \psi \right\rangle  & =& 
 \left\{ {\hbar \over i}\hat L_2
+ {\hbar \over 2}\cdot {{\sin \phi }\over {\sin \theta }} 
\right\} 
\ \left\langle \theta , \phi
\left\vert  \psi \right. \right\rangle ,\\
\left\langle \theta , \phi
\left\vert \hat {\bf S}_3 \right\vert \psi \right\rangle  & =&
{\hbar \over i}\hat L_3   
\ \left\langle \theta , \phi
\left\vert  \psi \right. \right\rangle .
\end{eqnarray}
These operators induce the
half-spin:
\begin{equation}
\left\vert  \psi_t \right\rangle 
=
c_+(t)
\left\vert +  \right\rangle +
c_-(t)
\left\vert -  \right\rangle ,
\end{equation}
where the eigen states have the following expression:
\begin{equation}
 \left\langle \theta , \phi ,\chi
\left\vert  +
\right. 
\right\rangle
=  {1 \over {\sqrt{ 2\pi } } } 
e^{{i  \over 2 } (\phi + \chi )}
\cos  {\theta  \over 2 }   \ \ , \ \ \ \
 \left\langle \theta , \phi ,\chi
\left\vert  -
\right. \right\rangle
=  {1 \over {\sqrt{ 2\pi } } } 
e^{-{i  \over 2 } (\phi - \chi ) }
\sin   {\theta  \over 2 }  .
\end{equation}
whose reduced version is
\begin{equation}
 \left\langle \theta , \phi 
\left\vert  +
\right. 
\right\rangle
=  {1 \over {\sqrt{ 2\pi } } } e^{-i s  } 
e^{i{\phi  \over 2 } }
\cos  {\theta  \over 2 }   \ \ , \ \ \ \
 \left\langle \theta , \phi 
\left\vert  -
\right. \right\rangle
=  {1 \over {\sqrt{ 2\pi } } }  e^{-i s  }
e^{-i {\phi  \over 2 }  }
\sin   {\theta  \over 2 }  .
\end{equation}
They satisfy
\begin{equation}
\left.
 \hat {\bf S}\cdot  \hat {\bf S}
\left\vert    \pm  \right.\right\rangle
=   {3\over 4}\hbar^2
\left\vert \pm \right\rangle  \ \ , \ \ \ \
\left.
 \hat {\bf S}_3
\left\vert   \pm   \right.\right\rangle
=   \pm {\hbar \over 2} 
\left\vert \pm \right\rangle .
\end{equation}
In addition,
we can introduce the increasing operator 
and the decreasing one $\hat {\bf S}_{\pm}=
\hat {\bf S}_{1} \pm i \hat {\bf S}_{2}$:
\begin{equation}
\left\langle \theta , \phi  ,\chi
\left\vert \hat {\bf S}_{\pm} \right\vert \psi \right\rangle 
=   \hbar e^{\pm i \phi }
\left\{ \pm {{\partial }\over {\partial \theta }}
+icot \theta {{\partial }\over {\partial \phi }}
-{{i}\over {sin \theta }}{{\partial }\over {\partial \chi }}
 \right\} 
\ \left\langle \theta , \phi ,\chi
\left\vert  \psi \right. \right\rangle ,
\end{equation}
which proves the following relations:
\begin{equation}
 \hat {\bf S}_{\pm}
\left\vert    \mp  \right\rangle
=  
\left\vert \pm \right\rangle  \ \ , \ \ \ \
\hat {\bf S}_{\pm}
\left\vert    \pm \right\rangle
=   0 .
\end{equation}
As in the usual expression \cite{Dirac}
originated by Pauli,
if ketvectors $\vert \pm \rangle $
are denoted as
\begin{equation}
\left\vert  +
\right\rangle  =\left( {\matrix{ 1\cr
0\cr
}} \right)  \ \ , \ \ \ \
\left\vert  -
\right\rangle  =\left( {\matrix{ 0\cr
1\cr
}} \right) ,
\end{equation}
then,
$  \hat {\bf S}_j =    {\hbar \over 2} \sigma _j $
for the Pauli matrices:
\begin{equation}
\sigma _1 =\left( {\matrix{0&1\cr
1&0\cr
}} \right)
  \ \ , \ \ \ \
\sigma _2 =\left( {\matrix{0&-i\cr
i&0\cr
}} \right) 
 \ \ and  \ \ \ \
\sigma _3 =\left( {\matrix{1&0\cr
0&-1\cr
}} \right)  ;
\end{equation}
\begin{equation}
\sigma _+ =\left( {\matrix{0&1\cr
0&0\cr
}} \right)
  \ \ and \ \ \ \
\sigma _- =\left( {\matrix{0&0\cr
1&0\cr
}} \right)  .
\end{equation}

A general  state 
of
the half-integer spin
of a particle
has the following expression:
\begin{equation}
\left\vert   \psi_t   \right\rangle
= 
\sum_{m =  -l-1}^{  l}
\ 
c_m^{l+1/2}(t)
\left\vert    l+1/2,m +1/2  \right\rangle ,
\end{equation}
where, for the normalization constant $N_{l+1/2}^{m+1/2}$,
\begin{eqnarray}
 \left\langle \theta , \phi \right. 
\left\vert l+1/2; m +1/2 \right\rangle
&=&   N_{l+1/2}^{m+1/2} \sqrt{{l+m+1}\over {2l+1}}
e^{-i s  } 
e^{i{\phi  \over 2 } }
\cos  {\theta  \over 2 } \  Y^{m}_l(\theta , \phi ) \\
& \ & \ \ \
 + N_{l+1/2}^{m+1/2}  \sqrt{{l-m}\over {2l+1}} e^{-i s  }
e^{-i {\phi  \over 2 }  }
\sin   {\theta  \over 2 } \
 Y^{m+1}_l(\theta , \phi ) ;
\end{eqnarray}
and the eigen states satisfy 
\begin{eqnarray}
\left.
 \hat {\bf S}\cdot  \hat {\bf S}
\left\vert    l+1/2,m +1/2    \right.\right\rangle
&=& \hbar^2 (l+1/2)(l+3/2) 
\left\vert  l+1/2; m +1/2 \right\rangle \ \ , \\
\left.
 \hat {\bf S}_3
\left\vert    l+1/2,m +1/2    \right.\right\rangle
&=&   \hbar (m+1/2) 
\left\vert  l+1/2; m +1/2 \right\rangle .
\end{eqnarray}

Let us assume the classical motion of a rigid rotor
has the following Hamiltonian:
\begin{equation}\label{half-spin-hamiltonian}
H =  I^{-1}S\cdot S
+ S\cdot B  .
\end{equation}
To elucidate
that Hamiltonian (\ref{half-spin-hamiltonian})
has no trouble in the operator-ordering problem,
we can introduce 
\begin{equation}\label{half-spin-hamiltonian}
H =   I^{-1}r^2 (p^2+C^2p^{\prime 2}) + x \cdot \left\{ p\times C p^{\prime } \right\}
+ x\cdot \left\{ p\times (B-C p^{\prime } ) \right\}  
,
\end{equation}
where the induced motion
preserve the following initial conditions:
\begin{equation}\label{half-spin-hamiltonian}
x\cdot p =0  \ \ and \ \  p^{\prime } = {\hbar \over 2} .
\end{equation}
For Hamiltonian (\ref{half-spin-hamiltonian}),
the infinitesimal generator  of motion is
equivalent to the following observable:
\begin{equation}
\hat {\bf H} =  I^{-1}
 {\hat {\bf S}\cdot \hat {\bf S}  } +
{1\over 2}
\left\{ \hat {\bf S}\cdot B +
B\cdot \hat {\bf S} \right\}  
,
\end{equation}
where
\begin{equation}
C={\hbar \over 2}\left( {x \over {2 \left( x^2 +y^2\right) }}  ,
{y \over {2 \left( x^2 +y^2\right) }} , 0\right) +x\times \nabla s .
\end{equation}
Now, we can investigate the internal structure of such
a
half-integer spin particle, an  quark
or lepton as an
electron or a constituted particle
as a nucleus,
which would have the following spin 
for the internal three-dimensional Euclid space:
\begin{equation}\label{internal spin}
S (x,p)=    x
\times \left( p + \nabla s \right)
 + {\hbar \over 2}\left( {x \over {2 \left( x^2 +y^2\right) }}  ,
{y \over {2 \left( x^2 +y^2\right) }} , 0\right) .
\end{equation}
Such an interpretation
of half-integer spin
allows us to describe the Dirac
equation as the equation of the motion
for the following Hamiltonian:
\begin{equation}
H (x, p,\alpha ,\beta )=   \alpha_1\beta \cdot  \left( p -  {e\over c}A \right) 
+mc^2  \alpha_3
- {e}A_0  ,
\end{equation}
where
$\alpha $ 
and $\beta $
are the internal spins
expressed as relation (\ref{internal spin}).
Since the obtained Hamiltonian
is also canonical as discussed in the previous subsection,
it has the following  infinitesimal generator:
\begin{equation}
\hat {\bf H} =  \left( \hat \gamma_ j
\left( \hat {\bf p}^j  -  {e\over c}A \right) 
+mc^2 \right) \hat \gamma_0
- {e}A_0 ,
\end{equation}
where $ \hat \gamma $ is the Dirac matrices.
In the same way,
the internal freedom like the isospins
of a particle
can be expressed as the invariance of motion,
if its Lie group is a subset
of the infinite-dimensional semidirect-product group
$S(M)$.
More detailed consideration
on the relativistic quantum mechanics
will be held elsewhere.

\section{CONCLUSION}
The present paper
proved that SbM and then protomechanics
deduces both classical mechanics and quantum mechanics
in its natural consequence,
and supported the rigid-body interpretation of a
half-integer spin.
The next paper \cite{SbMIII} will discuss 
the intimate relationships between the present theory
and the other quantization methods known in 
twentieth century;
and it will reveal that the new interpretation 
of the measurement process is
compatible with reality and causality.

\setcounter{equation}{0}
\renewcommand{\theequation}{A\arabic{equation}}
\section*{APPENDIX: INTEGRATION ON MANIFOLD}\label{on Manifold}

Let us here determine  the properties of the manifold $ M$
that is
 the three-dimensional
physical space
for the particle motion
in classical or quantum mechanics,
or the space of graded field variables
for the field motion
in classical or the quantum field theory.

Let  $ \left(  M,  
{\cal O}_M\right) $ be
a Hausdorff space
for the family ${\cal O}_M$ of its open subsets,
and also
a
N-dimensional oriented
$C^{\infty }$
manifold that is
modeled
by the N-dimensional Euclid space ${\bf R}^N$
and thus it
 has an atlas
$\left( U_{\alpha } ,
\varphi_{\alpha }
\right)_{\alpha \in \Lambda_M} $
(the set of a local chart of $M$)
for some  countable set $\Lambda_M$ such that\\
\begin{enumerate}
\item  $M= \bigcup_{\alpha \in \Lambda_M}  U_{\alpha } 
 $,
\item  $\varphi_{\alpha }: U_{\alpha } \to  V_{\alpha }$
is a $C^{\infty }$ diffeomorphism
for some $V_{\alpha }\subset {\bf R}^N$ and
\item  if $ U_{\alpha } \cap  U_{\beta } \neq \emptyset $,
then $ g^{\varphi }_{\alpha \beta }=  
\varphi_{\beta }\circ \varphi_{\alpha }^{-1}  :
 V_{\alpha } \cap  V_{\beta }  \to
 V_{\alpha } \cap  V_{\beta } 
$ is a $C^{\infty }$ diffeomorphism.
\end{enumerate}
The above definition would
be extended
to include that of the infinite-dimensional manifolds
called ILH-manifolds.
A ILH-manifold
that is modeled by the infinite-dimensional  Hilbert space
having an inverse-limit topology instead of ${\bf R}^N$
\cite{Omori}.
We will, however, concentrate ourselves on
the finite-dimensional cases for simplicity.
Let us further assume that
$M$ has no boundary
$\partial M = \emptyset $
for the smoothness of
the $C^{\infty }$ diffeomorphism group $D(M)$
over $M$, i.e.,
in order to consider the 
mechanics on a manifold $N$
that has the boundary $\partial N \neq \emptyset $,
we shall substitute the doubling of $N$ for $M$:
$M=   N\cup \partial   N \cup  N $.

Now, manifold
 $ M $ is
the
topological measure space 
 $  M = ( M , {\cal B}\left( {\cal O}_M \right)
,  {\it vol}) $
that
has the volume measure ${\it vol}$
for the topological $\sigma $-algebra
${\cal B}\left( {\cal O}_M \right) $.
For the Riemannian  manifold $M$,
the (psudo-)Riemannian structure induces
the volume measure ${\it vol}$.

Second, we assume that
the particle moves on manifold $M$ and
has
its internal freedom represented by
a oriented manifold
$F= (F, {\cal O}_F) $,
where ${\cal O}_F$ is the family of
open subsets of $F$.
Let $F= (F, {\cal B}\left( {\cal O}_F\right) ,
m_F) $ be the topological measure space 
with the invariant measure $m_F$
under the group transformation
$G_F$: $\tilde g_* m_F  =m_F $ for 
$\tilde g \in G_F$
where
$\tilde g_* m_F\left(
\tilde g(A)\right)  =m_F (A)$ for $A\in {\cal B}\left(
{\cal O}_F\right) $.
In this case, 
the  state of the particle 
can be
represented as a position  on
the 
locally trivial,
oriented
fiber bundle
$E  = ( E , M  , F , \pi ) $  
with fiber $F$
over $M $
with a canonical projection 
$\pi  :E  \to M $,
i.e., for every
$x \in M$,
there is an open neighborhood $  U (x) $ and a
$C^{\infty }$ diffeomorphism
$\phi_{U } : \pi^{-1}\left( U(x)\right) \to U(x) \times F $ such that
$ \pi = \pi_U \circ \phi_{U } $
for $\pi_U : U(x)\times F \to U (x) :(x, s) \to x $.
Let $G_F$ be
the structure group of fiber bundle $E$:
the mapping 
$\tilde g_{\alpha \beta }=\phi_{U_{\alpha }}
\circ\phi_{U_{\beta }}^{-1}:
U_{\alpha }\cap U_{\beta } \times
F \to U_{\alpha }\cap U_{\beta } \times
F $ satisfies 
$\tilde g_{\alpha \beta }(x,s)  \in G_F$
for $(x,s) \in U_{\alpha } \cap \ U_{\beta } \times F$
and
 the cocycle condition:
\begin{equation}\label{cocycle}
\tilde g_{\alpha \beta } ( x,s )
\cdot \tilde g_{\beta \gamma } ( x,s ) =
\tilde g_{\alpha \gamma } ( x,s )  
\ \ \ \ \ \ \ \
for \ \ ( x,s ) \in U_{\alpha }\cap U_{\beta }\cap 
U_{\gamma } \times F,
\end{equation}
where $\alpha , \ \beta , \ \gamma \in \Lambda_M$;
and
condition (\ref{cocycle}) includes the following relations:
\begin{equation}
\tilde g_{\alpha \alpha } (x,s)  = id. \ \
\ \  for \ \  x\in U_{\alpha }, \ \ \ and \ \ \ \
\tilde  g_{\alpha \beta } (x,s)  =\tilde  g_{\beta \alpha }(x,s) ^{-1} 
\ \ \ \ 
for \ \ (x,s) \in U_{\alpha }\cap U_{\beta }\times F.
\end{equation}
Thus, $\left( E ,
{\cal O}_E\right) $ is the Hausdorff space
for the family ${\cal O}_E$ of the open
subsets of $E$ such that
$ \tilde U \in {\cal O}_E $
satisfies $
\phi_{U_{\alpha } } \left(
\tilde U \right) =
U_{\alpha }  \times 
U_{\alpha }^{\prime } $
for some $U_{\alpha } $ ($\alpha \in \Lambda_M$) and 
$U_{\alpha }^{\prime } \in {\cal O}_F$.

Now, $\left( E ,{\cal B}\left(
{\cal O}_E\right) , m_E \right) $ becomes
the topological measure space
with the measure $m_E$ 
induced by the measures
${\it vol} $ and $m_F $ as follows.
For $A \in
{\cal B}\left( {\cal O}_E \right) $,
there exists the following disjoint union
corresponding to the covering $M=
\bigcup_{\alpha \in \Lambda_M } U_{\alpha } $
such that
\begin{enumerate}
\item $A = \bigcup_{
\alpha \in \Lambda_M }A_{\alpha } $ 
where \ $\pi \left( A_{\alpha }
\right)  \subset U_{\alpha }$, and
\item $A_{\alpha }\cap A_{\beta }
= \emptyset $ for $\alpha \neq \beta $.
\end{enumerate}

Thus, the measure $m_E $
can be defined as
\begin{equation}
m_E (A) = \sum_{\alpha \in \Lambda_M  }
\left( {\it vol} \otimes m_F \right)
\circ \phi_{U_{\alpha }} (A_{\alpha }) 
.
\end{equation}
Notice that the above definition of
$m_E$ is independent of the choice of $\left\{ 
A_{\alpha }\right\}_{\alpha \in \Lambda_M} $
such that $A = \bigcup_{
\alpha \in \Lambda_M }A_{\alpha } $ 
is a disjoint union
since $m_F$ is the invariant measure 
on $F$ for the group transformation of $G_F$.

Let us introduce the
space ${\cal M}\left(E \right) $ of all
the possible {\it probability Radon
measures} for
the particle positions on $E$ 
defined as follows:
\begin{enumerate}
\item every $\nu \in {\cal M}\left(E \right) $
is the linear
mapping $\nu : C^{\infty }(E)\oplus {\bf M} \to {\bf R}$
such that
$\nu (F ) < +\infty  $
for $F \in C^{\infty }(E)$, and
\item for every $\nu \in {\cal M}\left(E \right) $,
there exists a 
$\sigma$-additive positive measure $P$
such that
\begin{equation}
\nu  (F) = \int_{E} dP(y ) \left( F (y )  \right)
\end{equation}
and
 that $ P(M) =1 $, i.e.,
$\nu \left( 1 \right) = 1  $.
\end{enumerate}
For every $\nu \in {\cal M}\left(E \right)$,
the probability density
function
(PDF) $\rho \in L^1 \left(
E, {\cal B}({\cal O}_E)\right)
$ 
is 
the positive-definite, 
and satisfies
\begin{eqnarray}
\label{measure's relation 0}
\nu \left( F    \right)  
&=&\int_{E= \cup_{\alpha \in \Lambda_M}
A_{\alpha } } dm_E ( y )  \
\rho   ( y )  \left(
F  
( y  )   \right) \\
&=& \sum_{\alpha \in \Lambda_M}
\int_{ \phi_{U_{\alpha }}(A_{\alpha })} d{\it vol} (x)\
dm_F (\vartheta )  \
\rho \circ  \phi_{U_{\alpha }}^{-1} ( x , \vartheta )  \left(
F  \circ  \phi_{U_{\alpha }}^{-1} 
( x , \vartheta  )   \right)  ,
\end{eqnarray}
where $dP = dm_E \otimes \rho   $.

\end{document}